\documentclass[twoside]{article}

\usepackage[accepted]{aistats2023}
\usepackage{bbm}
\usepackage{graphicx}
\usepackage{xcolor}
\usepackage{bbm}
\usepackage{amsmath}
\usepackage{amssymb}
\usepackage{natbib}
\newtheorem{proposition}{Proposition}
\newtheorem{cor}{Corollary}
\newtheorem{as}{Assumption}
\newcommand{\IPM}{\text{IPM}}
\newcommand{\minus}{{-}} %remove spaces around minuses
\newcommand{\equal}{{=}}
\newcommand{\plus}{{+}}
\usepackage{hyperref}
\usepackage{subfigure}
\usepackage{caption}

\usepackage{wrapfig}
\usepackage{enumitem}
\setlist[enumerate]{noitemsep, nolistsep}

\newcommand{\indep}{\small {\raisebox{0.05em}{\rotatebox[origin=c]{90}{$\models$}}}}
\newcommand{\squish}[1]{{#1\parfillskip=0pt\par}}
\begin{document}

% If your paper is accepted and the title of your paper is very long,
% the style will print as headings an error message. Use the following
% command to supply a shorter title of your paper so that it can be
% used as headings.
%
%\runningtitle{I use this title instead because the last one was very long}

% If your paper is accepted and the number of authors is large, the
% style will print as headings an error message. Use the following
% command to supply a shorter version of the authors names so that
% they can be used as headings (for example, use only the surnames)
%
%\runningauthor{Surname 1, Surname 2, Surname 3, ...., Surname n}

\twocolumn[

\aistatstitle{Understanding the Impact of Competing Events on Heterogeneous Treatment Effect Estimation from Time-to-Event Data}

\aistatsauthor{Alicia Curth \And Mihaela van der Schaar}

\aistatsaddress{University of Cambridge \And  University of Cambridge, The Alan Turing Institute } ]

\begin{abstract}
We study the problem of inferring heterogeneous treatment effects (HTEs) from time-to-event data in the presence of \textit{competing} events. Albeit its great practical relevance, this problem has received little attention compared to its counterparts studying HTE estimation without time-to-event data or competing events.   We take an outcome modeling approach to estimating HTEs, and consider how and when existing \textit{prediction} models for time-to-event data can be used as plug-in estimators for potential outcomes. We then investigate whether competing events present new challenges for HTE estimation -- in addition to the standard confounding problem --, and find that, because there are multiple \textit{definitions} of causal effects in this setting --  namely total, direct and separable effects --, competing events \textit{can} act as an additional source of covariate shift depending on the desired treatment effect interpretation and associated estimand. We theoretically analyze and empirically illustrate when and how these challenges play a role when using generic machine learning prediction models for the estimation of HTEs.
\end{abstract}

\section{\uppercase{Introduction}} \squish{Competing events are ubiquitous in medical applications where the focus is on the time until occurrence of an adverse event due to a specific cause \citep{lim2010methods, lambert2010estimating}. Especially when patients have comorbidities, the effect of a treatment on an event of interest can only be assessed when taking into account the presence of risk due to a competing event. For example, when assessing the effectiveness of different cancer treatments for individual cancer patients one may have to consider how to take into account how an individual's risk for cardiovascular events changes due to treatment. This question, however, is far from straightforward: competing events -- which act as \textit{mediators} of the treatment on the outcome of interest --  give rise to multiple and different definitions of counterfactual risk that could be used depending on the policy or research question of interest, as recently formalized in \cite{young2020causal}. To see this, note that a treatment which causes a high number of cardiovascular events will automatically result in fewer events due to cancer -- which will appear as a protective (total) effect of treatment on the risk of events due to cancer, but may not be the desired interpretation of \textit{what makes a treatment effective} against an adverse outcome of interest. Instead, one could be interested in the direct effect of treatment on outcome (under elimination of competing events) or in the effect of the component in the treatment on outcome that acts only on the primary outcome \citep{young2020causal, stensrud2020separable}. }

\textbf{Related work.} Possibly because of this conceptual difficulty, heterogeneous treatment effect (HTE) estimation from time-to-event (TTE) data with competing events has received no attention from the machine learning (ML) literature yet. This stands in stark contrast with the ML literature on closely related problems -- (a) TTE prediction with competing events (sometimes also referred to as `competing risks') and (b) HTE estimation with other outcomes -- which has flourished in recent years. The literature on the former has adapted a variety of ML methods for risk prediction in the presence of competing events -- e.g. using Bayesian nonparametric methods in continuous time \citep{alaa2017competing, zhang2018nonparametric} and neural networks in discrete time \citep{lee2018deephit, wang2022survtrace}. The literature on the latter has focused on HTE estimation for binary or continuous outcomes, and has either provided model-agnostic strategies to estimate HTEs using \textit{any} ML method \citep{kunzel2019metalearners, nie2017quasi, kennedy2020optimal, curth2020} or adapted specific ML methods to correct for specific challenges of HTE estimation \citep{shalit2017estimating, curth2021inductive}. The largest stream of this literature has focused on analyzing and correcting confounding-induced \textit{covariate shift} \citep{shalit2017estimating, johansson2018learning, hassanpour2019counterfactual, assaad2020counterfactual}. Closest to our setting are two recent papers that have investigated covariate shift challenges inherent to HTE estimation for TTE data \textit{without} competing events: \cite{chapfuwa2021enabling} used generative models for counterfactual TTE analysis in continuous time and \cite{curth2021survite} used neural networks for discrete time analyses. An extended discussion of related work can be found in Appendix \ref{app:lit}.

\textbf{Outlook.} In this paper, we study heterogeneous treatment effect estimation in the presence of competing events through the causal framework recently established in \cite{young2020causal} and used therein to derive estimators for different \textit{average} treatment effects. We begin by considering how the ML toolbox developed for individualized risk prediction in the TTE setting with and without competing events could be used  for estimation of different HTEs through a potential outcome modeling approach, or conversely, consider what type of effects are implicitly targeted when different types of ML risk prediction algorithms are used as a basis for treatment decisions. As the ML literature on HTE estimation has focused on the presence of covariate shifts due to confounders, we then move to investigate which forms of covariate shift arise as we target different HTEs. We finally investigate and illustrate their effects empirically across simulation studies. Note that -- possibly unconventionally for this literature -- our focus here is not on designing or proposing a new method, but rather on \textit{understanding the unique challenges in a new and practically relevant problem} which is why we rely on simple existing methods to allow for clear insights. %

\squish{Our contributions are thus threefold: Conceptually, we study a new problem in the ML literature on HTE estimation and investigate how one could make use of the strong TTE estimation ML toolbox for solving it. Theoretically, we analyze covariate shift problems that arise therein. Empirically, we obtain insights into how estimation is affected in practice. Overall, we focus on \textit{understanding} the challenges underlying the problem, hoping that the insights that we provide will pave the way for future methodological work on this practically relevant problem.}

\section{\uppercase{Problem Setup}}
\begin{wrapfigure}{r}{0.45\columnwidth}
    \centering
    \includegraphics[width=0.45\columnwidth]{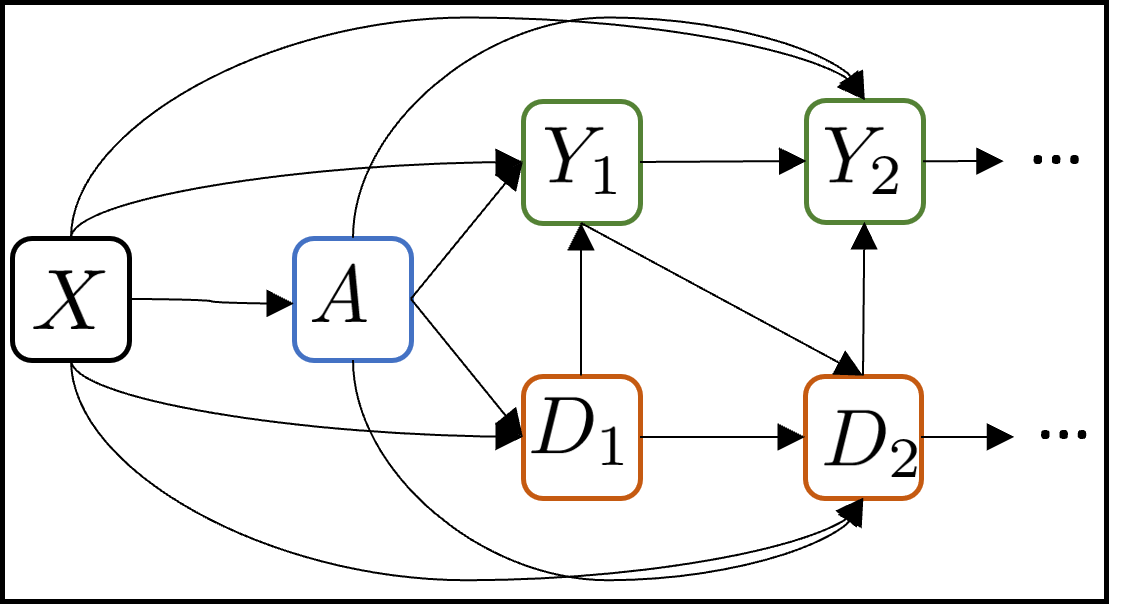}
    \vspace{-2em}
    \caption{Assumed Causal Graph} \vspace{-5mm}
 \label{simplegraph}
\end{wrapfigure}

{We adopt the setup of \citep{young2020causal, stensrud2020separable} in which patients are characterized by pre-treatment characteristics $X\!\!\in\!\!\mathcal{X}$, a binary treatment $A \!\!\in\!\! \{0, 1\}$ assigned at baseline, and $\{Y_k\}_{k \in \{1, \dots, K\}}$ and $\{D_k\}_{k \in \{1, \dots, K\}}$, binary indicators for whether the main event and competing event, respectively, have occurred \textit{by time period $k\!\leq\! K$}, where $K$ is the maximum time of follow-up. By convention, we assume that $D_k$ precedes $Y_k$, and that occurrence of either event precludes the other. Further, $\bar{V}_\kappa$ denotes the history $(V_0, \ldots, V_\kappa)$ of variable $V_k$ through interval $\kappa$. Fig. \ref{simplegraph} depicts the assumed underlying causal graph.  }

This data structure, which is in so-called long format, can equivalently be represented in short format of tuples $(X, A, T, E)$ where $T$ indicates the (discrete) time at which the event occurred (i.e. $T\equal\min k : Y_k \equal1 \lor D_k \equal 1$) and $E\!\in\!\{Y, D\}$ indicates its type (i.e. $E\equal Y \text{ if } Y_T\equal 1 \text{ else } D$).%\footnote{Note that we do not work with a latent failure time model as ... where each event e has failure time $T_e$ of which we observe only time $T=\min(T_Y, T_D)$, as this conceptually requires stronger assumptions on impossible events occuring \cite{prentice1978analysis}.} 

We will generally use long format as it uniquely allows to capture the sequential nature of the problem, but will sometimes use the short format when it simplifies notation.  We also assume \textit{no loss to follow-up} due to censoring (i.e. no patient drop-out) for simplicity, but discuss later in Sec. \ref{sec:covshift}  how censoring would play a role. 

Based on the assumed causal structure in Fig. \ref{simplegraph}, it is easy to see that we can model all risk functions of interest in the competing events literature -- e.g. the cause-specific cumulative incidence functions $ \mathbb{P}(T \leq k, E=Y|X=x, A=a)$ -- by relying on \textit{conditional hazard functions}. These are the hazard (probability) of the main event occurring given that there has been no event yet, i.e. 
\vspace*{-0.5em}
\begin{equation*}
  h_Y(k, x, a) =\mathbb{P}(Y_k\equal1|\bar{D}_k\equal\bar{Y}_{k-1}\equal0, X\equal x, A\equal a)  \vspace{-0.75em}
\end{equation*}

\vspace*{-0.5em}and, analogously, the hazard of the competing event occurring given event-free history 
\vspace*{-0.5em}
\begin{equation*}
    h_D(k, x, a)=\mathbb{P}(D_k\equal1|\bar{D}_{k-1}\equal\bar{Y}_{k-1}\equal0, X\equal x, A\equal a)
\end{equation*}

\vspace*{-0.5em}These can be used to model e.g. the cause-specific cumulative incidence function, or risk, of an event occurring by time $k$, as $\mathbb{P}(T \leq k, E\equal Y|X\equal x, A \equal a) \equal$
\vspace*{-0.4em}
\begin{equation}\label{basicriskeq}
\begin{split}
      \textstyle   \mathbb{P}(Y_k\equal1|X\equal x, A\equal a)    \equal\sum^k_{l=1} h_Y(l, x, a) \\ \textstyle
   \times \!\prod^{l-1}_{q=1} (1\minus h_Y(q, x, a)) (1\minus h_D(q, x, a))
 \end{split}
\end{equation}

\section{\uppercase{Defining and Estimating HTE Given competing events}}\label{sec:riskdefs}
\subsection{Preliminaries: Estimating HTEs from TTE data without competing events using prediction models}
To introduce the HTE estimation problem, counterfactuals and existing strategies for estimation, we begin with the simpler setting in which there are not competing events. Define the counterfactuals\footnote{Here, we use the term counterfactual following e.g. \cite{young2020causal, stensrud2020separable} exchangeably with the term potential outcome, which is different from \cite{pearl2009causality}'s usage of the term; within Pearl`s framework we only consider interventional quantities. Throughout, our counterfactuals/potential outcomes $Y^{a}$ or $Y^{a, \bar{d}}$ correspond to do-operations $do(A=a)$ or $do(A=a, \bar{D}_K=\bar{d})$, respectively.} (or potential outcomes) $Y^a_k$ for $a\!\in\!\{0,1\}$ and times $k\!\in\! \{1, \dots, K\}$ as the event indicator for the scenario in which a patient -- possibly countrary to fact -- has been \textit{assigned to} treatment $A\equal a$ at baseline, i.e. has been intervened on. Then, we can define heterogeneous treatment effects $\tau(x)$ as contrasts of functions of these potential outcomes: For example, if we are interested in \textit{differences in risk} of the event occuring by the end of study $K$, we have that for a patient with characteristics $X\equal x$, the treatment effect is
\begin{equation}
 \tau(x) = \mathbb{P}(Y_K^1=1|X=x) - \mathbb{P}(Y^0_{K}=1|X=x)
\end{equation}

\begin{figure*}[!tb]
\centering
\includegraphics[width=0.95\textwidth]{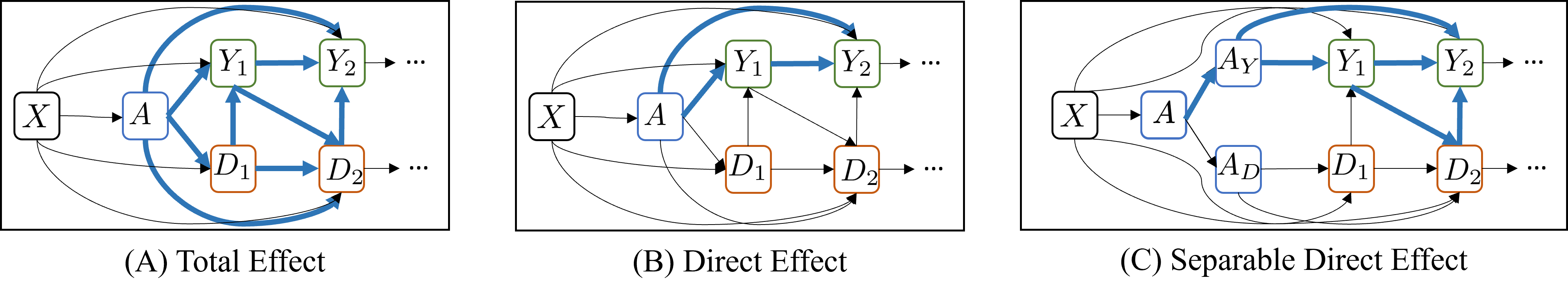}\vspace{-1.2em}
\caption{Illustration of the path of total, direct and separable direct effect of treatment $A$ onto $Y_2$.}\label{allpaths}\vspace{-1.5em}
\end{figure*}

Under the standard ignorability assumptions \citep{rosenbaum1983central} -- which ensure that $\bar{Y}_k^a \indep A | X$ for all $k$ and randomness in treatment assignment for all $x$ -- we have that $\mathbb{P}(Y^a_{k}=1|X=x)=\mathbb{P}(Y_{k}=1|X=x, A=a)$. Therefore a simple and popular way to \textit{estimate} HTEs is through outcome modeling: fitting a standard supervised learning model to predict $Y_K$ as 
\begin{equation*}
\hat{\mu}^a(x)=\hat{\mathbb{P}}(Y_{K}=1|X=x, A=a)
\end{equation*}
e.g. by appending $A$ to $X$ like a standard covariate,  or by fitting two separate prediction models, one on each treatment group (these two strategies are often referred to as S- and T-learner, respectively \citep{kunzel2019metalearners}). Then, the HTE can be estimated as $\hat{\tau}(x)=\hat{\mu}^1(x)-\hat{\mu}^0(x)$.

\subsection{Total effects: \textit{Using competing event prediction models for estimating HTEs in the presence of competing events allows to estimate total effects}} In a competing events setting, we can similarly define counterfactuals under intervention on treatment $Y^a_k$ (and analogously for the competing event resulting in $D^a_k$). At first glance, applying the same treatment effect estimation strategy discussed above to the setting with competing events seems appealing: one could use one of the recently proposed predictive ML models for the cause-specific cumulative incidence function (e.g. \cite{lee2018deephit}'s DeepHit or any other model that allows to estimate eq. (\ref{basicriskeq})) and  estimate treatment effect on risk as the difference 
\begin{equation*}
  \hat{\tau}(x) \equal \hat{\mathbb{P}}(Y_K\equal1|X\equal x, A\equal1)\minus\hat{\mathbb{P}}(Y_K\equal1|X\equal x, A\equal 0)
\end{equation*}
which, under similar ignorability assumptions, formalized below, \textit{is} a valid approach.

\begin{as}[Ignorability w.r.t. treatment]\label{ass:treat}  For each $k\in\{1, \ldots, K\}$, we have: (i) Exchangeability w.r.t. $A$: $Y^a_k, D^a_k\indep A|X$,  (ii) Positivity w.r.t. $A$: $\mathbb{P}(A=a|X=x)>0$ for $\forall x: \mathbb{P}(X=x)>0$ and $a\in \{0, 1\}$ and (iii) Consistency w.r.t. $A$: We observe the counterfactuals associated with the given treatment $A$, i.e. $Y_k=AY^1_k+ (1-A)Y^0_k$ and $D_k=AD^1_k+ (1-A)D^0_k$
\end{as}

 However, using competing event prediction models, which output cause-specific risks, in this way estimates a specific \textit{type} of treatment effect -- a total effect \citep{young2020causal} -- which may not always be the effect of most natural interest to an investigator. This is because probabilities outputted by cause-specific models depend not only on the occurrence of the primary event, but also on the competing event: Even when the treatment does not affect the primary event at all, it is possible that $\mathbb{P}(Y^1_K=1|X=x) \neq \mathbb{P}(Y^0_K=1|X=x)$ if treatment affects the competing event because this affects how many individuals are \textit{available} to experience the event (in other words, competing events act as \textit{mediators} in this context \citep{young2020causal}). Formally, this is because, as can be seen in eq. \ref{basicriskeq}, $\mathbb{P}(Y^a_K=1|X=x)$ depends on the conditional hazard of both the primary and the competing event -- thus even if $h_Y(k, x, a)$ is independent of $a$, the cause-specific risk may not be if $h_D(k, x, a)$ changes with treatment. Fig. \ref{allpaths}(A) illustrates the treatment effect's path associated with a total effect. To see why this may be undesirable, consider a cancer treatment that causes all patients in some subgroup in the treatment arm to experience heart failure but has no actual effect on their cancer-related events: in this case, the total effect would show that treatment \textit{reduces} the total risk of events due to cancer in this subgroup, but this is only true because no patients are \textit{available} to experience cancer-related events.  Note that this is not a problem with the identification of effects, but rather a feature when using \textit{cause-specific} risks to estimate effects. Thus, when using cause-specific risks and associated total effects for individualized treatment decision making one should be aware that this results in entanglement of different treatment effect pathways. 

\textit{Remark: Focusing on all-cause survival.} A simple way to overcome the competing events problem could be to \textit{combine} outcomes as $Y^{all}_k= Y_k \lor D_k$, i.e. letting go of the distinction of causes, and thus consider the \textit{overall} effect of treatment on all-cause survival. We do not consider this approach further here as this changes the outcome of interest, does not give competing events a special status and can thus be regarded a simple TTE analysis problem to be solved with e.g. the strategies discussed in \cite{chapfuwa2021enabling, curth2021survite}.

\subsection{Direct Effects: \textit{Using TTE models that treat competing events as censoring allows to estimate direct effects}}\label{sec:direff} Another type of counterfactual one could therefore be interested in estimating is $Y^{a, \bar{d}=0}_K$, where an \textit{additional intervention} is made and the competing event is eliminated: That is, an intervention that sets the entire history $\bar{D}_K$ to the deterministic value $\bar{d}=0$. Considering differences $\tau(x)=\mathbb{P}(Y^{1, \bar{d}=0}_K=1|X=x)-\mathbb{P}(Y^{0, \bar{d}=0}_K=1|X=x)$ then corresponds to \textit{direct} effects of treatment onto the event of interest \citep{young2020causal}. Fig. \ref{allpaths}(B) illustrates the treatment effect's path associated with a direct effect. This effect is identified under Assumption \ref{ass:treat} and an additional strong assumption:

\begin{as}[Ignorability w.r.t. competing event.]\label{ass:dir} For each $k\in\{1, \ldots, K\}$, we have: (i) Exchangeability: w.r.t. $D$:  $Y^a_k \indep D^a_k|X, \bar{Y}_{k-1}\!\!\!=\!\!\!\bar{D}_{k-1}\!\!\!=\!\!\!0, A\!\!\!=\!\!\!a$, (ii) Positivity w.r.t. $D$: $\mathbb{P}(D_k=0|X\!\!=\!\!x, \bar{Y}_{k-1}\!\!=\!\!\bar{D}_{k-1}\!\!=\!\!0, A\!\!=\!\!a)\!\!>\!\!0$ whenever $\mathbb{P}(X=x, \bar{Y}_{k-1}=\bar{D}_{k-1}=0, A=a)>0$, (iii) Consistency w.r.t. elimination of $D$: For an observation with $A=a$ and $\bar{D}_k=0$ we observe the corresponding counterfactual, i.e. $\bar{Y}_k=\bar{Y}_k^{a, \bar{d}=0}$.
\end{as}

Then it is also possible to estimate direct effects from observational data, as $\mathbb{P}(Y^{a, \bar{d}=0}_K\equal1|X=x)=\mathbb{P}(Y_K\equal1|\bar{D}_K\equal0, A\equal a, X\equal x)$, using the formula
\begin{equation*}
\mathbb{P}(Y^{a, \bar{d}=0}_K\equal1|X\equal x) = \sum^K_{l=1} h_Y(l, x, a) \prod^{l-1}_{q=1} (1\minus h_Y(q, x, a))
\end{equation*}
where, relative to the total risk, the dependence on $h_D(l, x, a)$ has been removed. That is, the direct risk treats competing events like a \textit{source of independent censoring} \citep{young2020causal}. This direct effect can therefore also be estimated using off-the-shelf ML methods using outcome modeling. In this case, however, one would no longer need to model the competing event as a separate cause, but instead treat it as a \textit{censoring event} and use single survival models for $Y_K$ only as e.g. \cite{curth2021survite} or simply use only the cause-specific hazard functions $h_Y(l, x, a)$ if they are available from a competing events model.

Note that direct effects not only require stronger identifying assumptions than total effects, but the presence of the intervention $do(\bar{D}_K=0)$ also introduces a \textit{conceptual challenge}: such a hypothetical intervention may not always be feasible \citep{stensrud2020separable} -- e.g. an intervention eliminating \textit{all} risk of cardiovascular events in cancer trial could be somewhat hard to conceptualize.

\subsection{Separable Effects: \textit{Estimating path-specific (separable) effects requires access to hazard estimators}}
A final alternative effect definition, presenting fewer conceptual challenges than direct effects, was recently proposed in \cite{stensrud2020separable}: \textit{separable} direct and indirect effects are path-specific effects that assume $A$ conceptually consists of components $A_Y$ and $A_D$, which affect only the primary event $Y_k$ and the competing event $D_k$, respectively. While we have that $A\equal A_Y\equal A_D$ in the observed data, we could hypothesize an intervention that sets $A_D$ and $A_Y$ to separate values. This could be plausible if a treatment may consist of different active components with different biological functions that could be deactivated in the future \citep{stensrud2020separable} -- allowing to define counterfactuals $Y^{a_Y,a_D}_k$ (and $D^{a_Y,a_D}_k$) which can be used to e.g. investigate separable direct effects on risk $\mathbb{E}[Y^{1, a_D}_K - Y^{0, a_D}_K|X\equal x]$ and separable indirect effects on risk $\mathbb{E}[Y^{a_Y, 1}_K - Y^{a_Y, 0}_K|X\equal x]$. Fig. \ref{allpaths}(C) illustrates the treatment effect's path associated with a separable direct effect.

Risk under separable treatments can be estimated from observed data (where treatment was not separated) under Assumption \ref{ass:treat} and additional Assumption \ref{ass:dismissible}, which, as discussed in Appendix \ref{app:tech}, has  implications similar to Assumption \ref{ass:dir} needed for direct effect estimation.
\begin{as}[Identification w.r.t. separable treatment.]\label{ass:dismissible}  For each $k\in\{1, \ldots, K\}$, we have: (i) Dismissible components: 
\begin{equation*}
\begin{split}
\mathbb{P}(Y_k^{a_Y, a_D=1}\equal 1|\bar{Y}_{k-1}^{a_Y, a_D=1}\equal 0, \bar{D}_k^{a_Y, a_D=1}\equal 0, X\equal x)=\\ \mathbb{P}(Y_k^{a_Y, a_D=0}\equal 1|\bar{Y}_{k-1}^{a_Y, a_D=0}\equal 0, \bar{D}_k^{a_Y, a_D=0}\equal 0, X\equal x)    
\end{split}
\end{equation*} 
and a similar condition equalizing the conditional hazards of $D_k^{a_y=0, a_D}$ and  $D_k^{a_y=1, a_D}$ (see Appendix \ref{app:tech}), (ii) positivity w.r.t. $A$ (in surviving population): $\mathbb{P}(A=a|\bar{D}_k=\bar{Y}_k=0, X=x)>0$ whenever $\mathbb{P}(\bar{D}_k=\bar{Y}_k=0, X=x)>0$ for $a\in\{0,1\}$, (iii) Consistency: For an observation with $A=a$, we observe the corresponding counterfactuals, i.e. $Y_k=Y^{a, a}_k$ and $D_k=D^{a, a}_k$.
\end{as}

Then, risk under separable components can be estimated as $ \mathbb{P}(Y^{a_Y, a_D}_K = 1|X=x)=$
\begin{equation*}\label{sep_id}
 \sum^K_{l=1} h_Y(l, x, a_Y) \prod^{l-1}_{q=1} (1\minus h_Y(q, x, a_Y))(1\minus h_D(q, x, a_D))
\end{equation*}
which differs from the identification formula for total risk in that it evaluates the hazard of the competing event under treatment $a_D \neq a_Y$. As no current ML prediction models target such separable treatment paths, most TTE models cannot directly be used by e.g. including treatment as a standard covariate and simply issuing predictions; yet any TTE model from which conditional hazard estimates $\hat{h}_D(k, x, a)$ and $\hat{h}_Y(k, x, a)$ can be extracted can be used to estimate separable risk through computation of the formula above instead of issuing its standard predictions.

\textit{Remark: Implications for the use of TTE prediction models for treatment decision making.} Some existing work proposing ML TTE competing events prediction methods use the example of making treatment plans as motivation for their method (e.g. \citet{alaa2017competing}). We therefore wish to reemphasize that, through our discussion in this section, it becomes clear that the use of different types of TTE prediction methods implicitly means considering different types of effects in this context. Assuming that the necessary identifying assumptions hold, when TTE prediction models that explicitly model competing events (e.g. \citet{alaa2017competing, lee2018deephit}) are used to inform treatment plans, this implicitly corresponds to consideration of total effects, while approaches treating competing events as \textit{censoring} events implicitly lead to consideration of direct effects. As discussed above, separable effects are generally not implicitly a by-product of generic TTE prediction methods.

\section{\uppercase{Understanding Covariate shifts due to competing events}}
Assuming that all identifying assumptions described above\footnote{Appendix \ref{app:tech} contains extended discussions of assumptions.} hold, all different types of treatment effects can be estimated from observed data -- yet not without further challenges. Because the data available for training follows an observational distribution, while the target quantities are defined with respect to interventions, \textit{covariate shift} arises. Covariate shift arising due to treatment selection on observables (confounding) has been studied in detail in the ML literature on HTE estimation with standard (binary/continuous) outcomes since \cite{johansson2016learning, shalit2017estimating}. More recently, \cite{chapfuwa2021enabling} tackled only confounding-induced covariate shift in the context of survival outcomes, and \cite{curth2021survite} showed that censoring acts as an additional source of covariate shift in the TTE setting.

In this section, we take a closer look at how covariate shift can arise in the TTE setting with competing events when learning treatment-specific hazard functions from observational data. We show that, because the different effects are defined with respect to different interventions,  competing events \textit{can} act as an additional source of covariate shift depending on the chosen treatment effect of interest.

\subsection{Learning treatment-specific hazard functions}
Here, we focus on learning hazard functions because they can be (a) used to compute all treatment effects defined in the previous section, and (b) easily estimated using off-the-shelf ML methods simply by restricting the training set -- and can thus be analyzed like a standard supervised learning problem. The simplest and most flexible problem formulation, which we focus on here, imposes no assumption on how hazards evolve over time (e.g. no proportional hazards assumption) or on how treatment affects outcome, by fitting a separate model for each conditional hazard, giving an estimator for each time-step by treatment group by event type (i.e. $K\! \times \!2 \!\times\! 2$ estimators in total).

To do so, inspired by the approach described in e.g. \cite{stitelman2010collaborative, curth2021survite} in the standard TTE setting, we simply separate the observed data $\mathcal{D}_{obs}=\{(X_i, A_i, \bar{Y}_{K, i}, \bar{D}_{K, i} \}^n_{i=1}$ by treatment group, and then, for each time step $k$, first fit a classification model for outcome $D_k$ and then fit a classification model for outcome $Y_k$, by using only the patients still at risk of the events: That is, for each time-step $k$, we estimate $\hat{h}_D(k, x, a)\!\! =\!\! \hat{\mathbb{P}}(D_k\!\!=\!\!1|\bar{D}_{k-1}\equal \bar{Y}_{k-1}\equal 0,X\equal x, A\equal a)$ by solving a standard classification problem with input-output tuples $\mathcal{D}^D_{a, k}\!\!=\!\!\{(X_i, D_{k, i})\}_{i \in \mathcal{I}_D(k, a)}$ where $\mathcal{I}_D(k, a)\!\!=\!\!\{i \in [n]: \bar{D}_{k-1, i}\!\!=\!\!\bar{Y}_{k-1,i}\!\!=\!\!0, A_i\!\!=\!\!a\}$ is the competing at-risk set at time-step $k$ with $n^D_{k, a}\!\!=\!\!|\mathcal{I}_D(k, a)|$, and estimate $\hat{h}_Y(k, x, a)\!\!=\!\!\hat{\mathbb{P}}(Y_k\!\!=\!\!1|\bar{D}_{k, }\equal \bar{Y}_{k-1, }\equal 0,X\equal x, A \equal a)$ by fitting a classification model for input-output tuples $\mathcal{D}^Y_{a, k}\!\!=\!\!\{(X_i, Y_{k, i})\}_{i \in \mathcal{I}_Y(k, a)}$ using patients remaining in the main at-risk set $\mathcal{I}_Y(k, a)\!\!=\!\!\{i \in [n]: \bar{Y}_{k-1, i}\!\!=\!\!\bar{D}_{k, i}\!\!=\!\!0, A_i\!\!=\!\!a\}$, with $n^Y_{k, a}\!\!=\!\!|\mathcal{I}_Y(k, a)|$, at time-step $k$. 

\subsection{How does covariate shift arise?}\label{sec:covshift}
When fitting the conditional hazard for the main event\footnote{From here on, we focus on our discussion on the main event, but analogous derivations can be made when effects on the competing event are of interest, where the observational distribution is $\mathbb{P}(X=x|\bar{Y}_{k-1}\equal D_{k-1}\equal 0, A\equal a)$.} using empirical risk minimization (ERM)  for the classification approach described above, the hazard estimator is
\begin{equation*}
    \hat{h}^Y(k, x, a) \in \arg \min_{h \in \mathcal{H}} \hat{R}^{obs}_{a, k}(h) 
\end{equation*}
where $\textstyle \hat{R}^{obs}_{a, k}(h)\equal\sum_{i \in \mathcal{I}_Y(a, k)}\ell(Y_{k, i}, h(X_i))$ is the empirical version of  the risk $R^{obs}_{a, k}(h) = \mathbb{E}_{X \sim p^{obs}_{a, k},  Y_k \sim h_Y(k,x,a) }[\ell(Y_k, h(X))]$, $\mathcal{H}$ denotes the hypothesis class under investigation, $\ell$ is some loss function and $\mathbb{P}^{obs}_{a, k}$ refers to the \textit{observational} at-risk covariate distribution, i.e. $\textstyle \mathbb{P}^{obs}_{a, k}(X=x) = \mathbb{P}(X\equal x|\bar{Y}_{k-1}\equal \bar{D}_k\equal 0, A\equal a)$
\begin{equation}
\begin{split}
\textstyle\propto \mathbb{P}(X
\equal x)\mathbb{P}(A\equal a|X=x) (1\minus h_D(k, x, a))\\ \textstyle\times \prod^{k-1}_{l=1} (1\minus h_D(l, x, a))(1\minus h_Y(l, x, a))
\end{split}
\end{equation} 

Our target distribution, however, is not the observational distribution: because the treatment effects under consideration are associated with different types of interventions, the target covariate distribution corresponds to an interventional distribution $\mathbb{P}^{int}_{a, k}$, giving rise to the (hypothetical) interventional risk $R^{int}_{a, k}(h) = \mathbb{E}_{X \sim p^{int}_{a, k},  Y_k \sim h_Y(k,x,a) }[\ell(Y_k, h(X))]$. This mismatch between observed and target distribution is known as \textit{covariate shift} and has as a consequence that the learnt function is not necessarily optimal as it is possible that $\arg \min_{h \in \mathcal{H}} {R}^{int}_{a, k}(h) \neq \arg \min_{h \in \mathcal{H}} {R}^{obs}_{a, k}(h)$. It is well-known that such a mismatch can be addressed by relying on so-called \textit{importance weights} $w^*(x) \propto \frac{\mathbb{P}^{int}_{a, k}(X=x)}{\mathbb{P}^{obs}_{a, k}(X=x)}$ to give $\hat{R}^{w, obs}_{a, k}(h)\equal\sum_{i \in \mathcal{I}_Y(a, k)}w^{*}(X_i)\ell(Y_{k, i}, h(X_i))$ which is unbiased for the interventional risk. Below, we discuss the type of covariate shift and associated importance weights arising when conceptualizing the interventions for the three effects under investigation:

\textbullet{} \textbf{Total effect:} The total effect requires only an intervention on treatment  $do(A=a)$, thus, as can be read off from the causal graph in Fig. \ref{simplegraph}, the interventional at-risk distribution $\mathbb{P}^{do(A=a)}_{a, k}(X=x)$ is proportional to $\mathbb{P}(X\equal x)(1-h_D(k, x, a))\prod^{k-1}_{l=1} (1-h_D(l, x, a))(1-h_Y(l, x, a))$, which differs from the observational distribution only in the treatment assignment factor $\mathbb{P}(A\equal a|X\equal x)$. That is, if treatment was assigned completely at random as in a randomized trial, the two distributions $\mathbb{P}^{obs}_{a, k}$ and $\mathbb{P}^{int}_{a, k}$ would be the same. The covariate shift arising when estimating total effects is thus the same shift arising in the standard treatment effect setting considered in e.g. \cite{shalit2017estimating}, leading to time-independent importance weights $w^{*, do(A=a)}_{a, k}(x) \propto P(A\equal a|X\equal x)^{-1}$.

\textbullet{} \textbf{Direct effect:} The direct effect requires intervention $do(A\equal a, \bar{D}_K\equal 0)$, which, in addition to treatment assignment bias, also removes all competing events. Therefore, as can be read off from Fig. \ref{simplegraph}, the interventional at-risk distribution $\mathbb{P}^{do(A=a, \bar{D}_K=0)}_{a, k}(X=x) \propto \mathbb{P}(X\equal x)\prod^{k-1}_{l=1} (1\minus h_Y(l, x, a))$. This differs from the observational distribution in both the absence of the treatment assignment factor \textit{and} removes all effects of competing event on the population composition through removal of the factor $\prod^{k}_{l=1} (1\minus h_D(l, x, a))$; the latter results in covariate shift only if the risk of the competing event is dependent on covariates. As discussed in Section \ref{sec:direff}, the competing event is effectively treated as a censoring event here and the shift is thus equivalent to the censoring-induced shift discussed in \cite{curth2021survite}. The importance weights needed to correct for this shift would thus be  $w^{*, do(A\equal a, \bar{D}_K\equal 0)}_{a, k}(x) \!\!\propto\!\!\left[P(A\equal a|X\equal x )\prod^{k}_{l=1} (1\minus h_D(l, x, a))\right]^{-1}$.

\textbullet{} \textbf{Separable effects:} Finally, separable effects require separate interventions on both treatment components $do(A_Y\equal a_Y, A_D\equal a_D)$, thus the interventional distribution $\mathbb{P}^{do(A_Y\equal a_Y, A_D\equal a_D)}_{a, k}(X=x)$, which is identified due to the dismissible component condition (see Appendix \ref{app:tech}), is proportional to
\vspace{-1em}
    \begin{equation*}\begin{split}
        \mathbb{P}(X \equal x)
       (1 \minus h_D(k, x, a_D))\\ \times \prod^{k-1}_{l=1}  (1\minus h_D(l, x, a_D))(1\minus h_Y(l, x, a_Y))
       \end{split}
    \end{equation*}
 This differs from the observational distribution in the treatment assignment factor and in that it has $\prod^{k}_{l=1} (1- h_D(l, x, a_D))$ instead of $\prod^{k}_{l=1}(1- h_D(l, x, a_Y))$. Note that the latter results in covariate shift in the at-risk distribution only if \textit{the effect of treatment on the competing event} is dependent on covariates. The importance weights needed to correct for this shift would thus be  $w^{*,do(A_Y\equal a_Y, A_D\equal a_D)}_{a, k}(x) \propto \left[P(A=a|X=x)\frac{\prod^{k}_{l=1} h_D(l, x, a_Y)}{\prod^{k}_{l=1} h_D(l, x, a_D)}\right]^{-1}$.

\textit{Remark: How would the addition of censoring play a role?} As we noted above, direct effect estimation essentially corresponds to TTE estimation with ignorable censoring -- as one would usually `switch off' censoring for a treatment effect analysis \citep{young2020causal}. Thus, if we were to add an additional event $C_k$ to our setting to allow censoring (patient drop-out), this would (a) require additional identifying assumptions similar to assumption \ref{ass:dir}, e.g. $Y^a_k \indep C^a_k|X$, and (b) lead to the same covariate shift issues as a competing event in direct effect estimation setting -- any observational distribution would thus gain an additional factor $\prod^k_{l=1}\mathbb{P}(C_l=0|\bar{D}_{l-1}\!=\!\bar{Y}_{l-1}\!=\!\bar{C}_{l-1}\!=\!0, X\!=\!x, A\!=\!a)$, which should then be (inversely) multiplied with any importance weight.

%From the above we once more note that total effect is the \textit{standard} treatment effect problem in that the only source of covariate shift is treatment assignment (i.e. there is no covariate shift in clinical trial setting), while the other two are not: even in a clinical trial, covariates shift and the at-risk population is different from the interventional distribution of interest. 

\subsection{How does this covariate shift affect estimation of hazards from observational data?}
To analyze how the learning of hazard functions will be impacted by the multitude of sources of covariate shift we outlined above, we can apply well-known results from the literature on domain adaptation and importance weighting to our problem. Below, we adapt the bound of \cite{cortes2010learning} to our setting; refer to Appendix \ref{app:tech} for proofs.

\begin{proposition}\label{prop:full} Given timestep $k$ and treatment $a$, for a loss function $\ell_h \in [0, 1]$ of any hypothesis $h \in \mathcal{H}$, such that $d=\text{Pdim}(\{\ell_h: h \in \mathcal{H}\})$ (where Pdim is the pseudo-dimension) and $\ell_h \in \mathcal{L}$, where $\mathcal{L}$ is a space of pointwise loss functions, and a weighting function $w(x)$ with $\mathbb{E}[w(X)]=1$, with probability $1 - \delta$ over at-risk sample $\mathcal{D}^Y_{a, k}$ with empirical distribution $\hat{p}^{obs}_{a, k}$, we have 
\vspace{-0.9em}
\begin{equation*}
\begin{split}
\textstyle R^{int}_{a, k}(h) - \hat{R}^{w, obs}_{a, k}(h) \leq \left |\mathbb{E}_{p^{obs}_{a, k}}[(w^*_{a, k}(x) - w(x))\ell_h(x)]\right| \\ + 
\frac{\max(\sqrt{\mathbb{E}_{p^{obs}_{a, k}}[w^2(x)l^2_h(x)]}, \sqrt{\mathbb{E}_{\hat{p}^{obs}_{a, k}}[w^2(x)l^2_h(x)]})}{n^Y_{a, k}{}^{3/8}} \mathcal{C}^\mathcal{H}_{n^Y_{a, k}}
\end{split}
\end{equation*}
with $\textstyle\mathcal{C}^\mathcal{H}_{n^Y_{a, k}} = 2^{5/4}(d \log \frac{e2n^Y_{a, k}}{d} + \log \frac{4}{\delta})^{3/8}$ due to \cite{cortes2010learning} (Theorem 4).
\end{proposition}

The proposition above, broken down further in two corollaries below by incorporating ideas from \citet{johansson2018learning}
and \citet{maia2022effective}, tells us multiple things about the difficulty of the hazard estimation problem and the consequences of the arising covariate shifts. A first observation from the general statement above is that, unsurprisingly, the at-risk sample size $n^Y_{a,k}$ determines the speed of learning. This sample size naturally decreases in $k$ through the occurance of main events even in the absence of competing events, but decreases further as a function of the frequency of competing events, meaning that $n^{Y,obs}_{a, k} \leq n^{Y, int}_{a, k}$ for hypothetical datasets of size $n$ created using the interventions corresponding to direct (or separable) effects, where the inequality is strict if competing events exist (or if treatment influences the rate of competing events occurring).   

\begin{cor}[Perfect importance weights]\label{cor:weights}
For $w(x)\equal w^{*, int}_{a, k}(x)$, we have
\begin{equation*}
     R^{int}_{a, k}(h) \minus \hat{R}^{w^*, obs}_{a, k}(h) \!\!\leq\!\! \frac{1}{\sqrt{ESS^{*}_{rel}(p_{a,k}^{int}, p_{a,k}^{obs})} n^Y_{a,k}{}^{3/8}} \mathcal{C}^\mathcal{H}_{n^Y_{a, k}}
\end{equation*}
where $ESS^{*}_{rel}=\exp_2(D_2(p^{int}_{a,k}||p^{obs}_{a,k}))$ is the expected relative effective sample size, with $D_2(p||q)= log_2 \mathbb{E}_{x \sim p} \left[\frac{p(x)}{q(x)}\right]$ the R\a'enyi divergence of order 2.
\end{cor}
In Corollary \ref{cor:weights}, we see that for the special case of perfect importance weights the first term of the RHS in Proposition \ref{prop:full} is zero and the speed of learning is slowed down if the relative effective sample size $ESS^{*}_{rel}\!<\!1$ -- i.e. whenever interventional and observational covariate distributions differ. To gain further intuition,  note that \cite{maia2022effective} show that for self-normalized importance weights $\bar{w}^*_i$, we can approximate  $ESS^{*}_{rel}$ from data as $\frac{1}{n\sum_{i \in \mathcal{I}}\bar{w}^{*,2}_i} \rightarrow ESS^{*}_{rel}$ a.s. as $|\mathcal{I}| \rightarrow \infty$, which is lowest when all weights are equal. Time-dependent importance weights $\frac{p_{a,k}^{int}}{p_{a,k}^{obs}}$ may generally be more variable as $k$ increases, e.g. if $h_D(l, x, a)$ is constant in $l$, we have $({\prod^{k}_{l=1} h_D(l, x, a_D)})^{-1}=( h_D(l, x, a_D))^{-k}$, meaning that the covariate shift problem could be exacerbated over time for the direct and separable effects. However, as the density ratio is integrated w.r.t. $p^{int}_{a,k}$ in the R\a'enyi divergence, this effect can subside if the overall at-risk probability is small, as we also demonstrate empirically below.  

\begin{cor}[Standard supervised learning (unweighted)]\label{cor:noweight}
For $w(x)=1$, we have 
\begin{equation*}
    R^{int}_{a, k}(h) \minus \hat{R}^{obs}_{a, k}(h) \!\!\leq\!\! C_\mathcal{L} IPM_{\mathcal{L}}(p^{int}_{a,k}, p^{obs}_{a,k}) + n^Y_{a, k}{}^{-3/8} \mathcal{C}^\mathcal{H}_{n_{a, k}}
\end{equation*}
where $\IPM_\mathcal{G}(p, q) = \sup_{g\in \mathcal{G}}\left|\int g(x) (p(x) - q(x))dx\right|$ is an integral probability metric and $C_\mathcal{L}\!>\!0$ is s.t. $\frac{\ell}{C_\mathcal{L}}\in \mathcal{L}$.
\end{cor}

\begin{figure*}[t]
    \centering
    \includegraphics[width=0.9\textwidth]{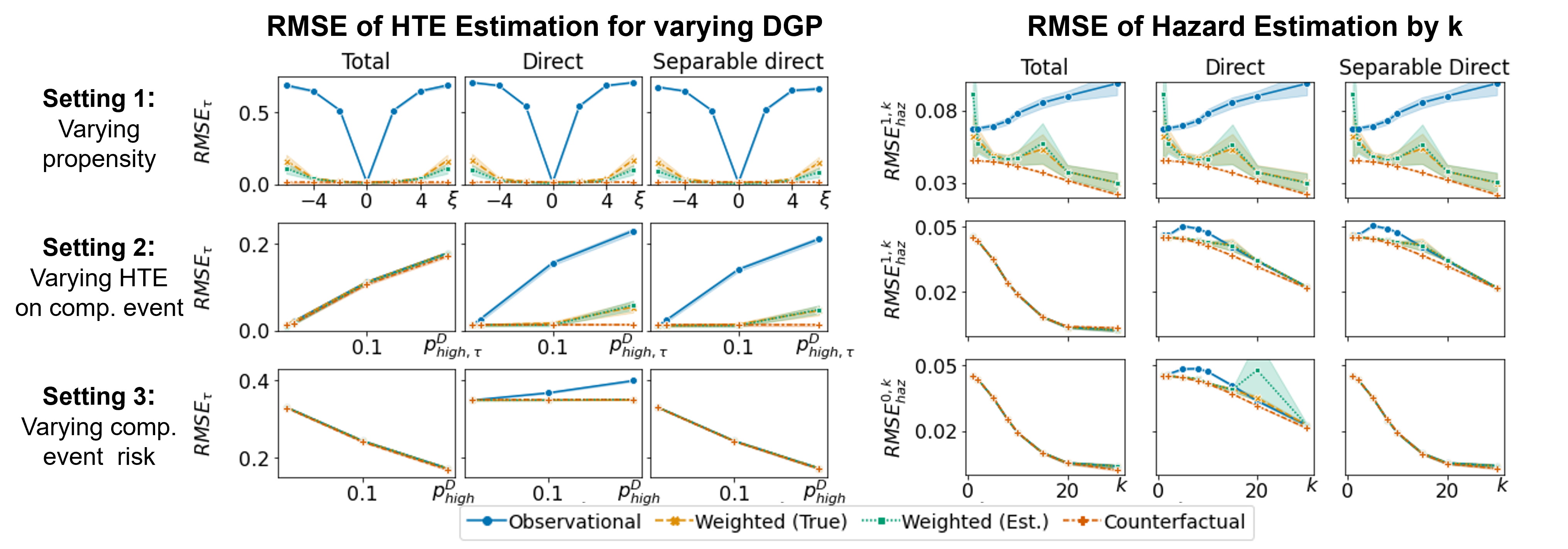}\vspace{-1em}
    \caption{Estimation performance in $\textstyle RMSE_\tau$ as parameters of the DGP vary (left) and  $\textstyle RMSE^{a,k}_{haz}$ over time $k$ (right), for the 3 effects (columns) across 3 settings (rows). For $\textstyle RMSE^{a,k}_{haz}$, each DGP`s varying parameter is fixed its highest value.  }\vspace{-1em}
    \label{fig:conf_res}
\end{figure*}

Finally, we consider the special case of \textit{no} weighting in Corollary \ref{cor:noweight}. Here, $ESS^{*}_{rel}\!=\!1$ due to constant weights and the first term of the RHS of \ref{prop:full} is bounded by an IPM-term -- which does \textit{not} decrease as the sample size grows\footnote{A large proportion of the ML HTE literature has therefore focused on learning representations that minimize an empirical estimate of the IPM-term \cite{shalit2017estimating}. This is easily possible in the standard setting as $\mathbb{P}^{int}$ is observed in the marginal covariate distribution $\mathbb{P}(X\!\!=\!\!x)$ and can hence be used to approximate the IPM term -- however, in our setting, the interventional distribution is an unobserved at-risk distribution that differs per time-step (i.e. $\mathbb{P}^{int}_{a, k}$ is not the marginal $\mathbb{P}(X\!\!=\!\!x)$ except for at $k\!=\!1$), giving no straightforward analogue to this approach.}\label{footnoteimp}, reflecting that standard ERM on the observational data may never recover the best interventional solution. As a consequence, $ R^{int}_{a, k}(h) \minus \hat{R}^{obs}_{a, k}(h)$ may never vanish -- depending also on how rich the underlying hypothesis class is. This is well-known to be a problem for misspecified parametric models\footnote{For correctly specified parametric  models and likelihood loss, we always have $\arg \min_{h \in \mathcal{H}} {R}^{obs}_{a, k}(h) = \arg \min_{h \in \mathcal{H}} {R}^{int}_{a, k}(h)$.} \citep{sugiyama2007covariate}, while, when using rich hypotheses classes through flexible nonparametric or deep methods, one generally does not have to trade off model performance in different regions of the covariate space, meaning that, given sufficient data, importance weighting would not be expected to make a difference \citep{byrd2019effect}.

\section{\uppercase{Experiments}}
Finally, we empirically investigate whether, when and how the different shifts play a role when learning hazard functions with the purpose of estimating the different HTEs. As is common practice in the HTE literature \citep{curth2021doing}, we have to rely on \textit{simulated} data because counterfactuals are not available in real data, meaning that real datasets provide \textit{no ground truth} for evaluating methods. While the standard HTE estimation problem in absence of competing events is only missing a single counterfactual (with respect to interventions on treatment assignment), the problem is exacerbated in our setting where additional (unobservable) interventions on competing events would be required to create ground truth targets for evaluation. In addition to the fully synthetic and highly stylized experiments considered below, we present additional results from a semi-synthetic setup using the real Twins dataset \cite{louizos2017causal} in Appendix \ref{app:twins}, leading to similar insights as the results presented below.

\textbf{An illustrative DGP.}  Because there are many different forces at play which we wish to disentangle, we focus on a simple setup here that allows us to highlight important problem features. We assume that individuals are characterized by $x\equal(x_1, x_2)$,  two binary risk factors $X_1 \!\sim\! \mathcal{B}(0.5)$ and $X_2 \sim \mathcal{B}(0.5 \minus \rho (1 \minus 2X_1))$ that may be correlated; unless indicated otherwise we set $\rho\equal 0.35$. We assume a very simple hazard for both outcomes $E \!\in\! \{Y, D\}$:
\begin{equation}
    h_E(k, x, a) =  \begin{cases}
    p^E_{low} + a p^E_{low, \tau} \text{ if } x_{S_E}=0\\
     p^E_{high} + a p^E_{high, \tau} \text{ if } x_{S_E}=1\\
    \end{cases}
\end{equation}
with $0 \!< \!p^E_{\cdot} \plus p^E_{\cdot, \tau}\! \leq\! 1$ for both settings.  This model is constant over time and depends only on the covariate $x_{S_E}$, where $S_E$ is the index of the support covariate for the event model. We let $S_Y=S_D=1$, and, unless otherwise indicated, set $p^Y_{low}\equal p^D_{low}\equal p^D_{high}=0.01$, and create a high primary outcome risk group with $p^Y_{high}\equal 0.1$ and assume no treatment effect $p^E_{high, \tau} \equal p^E_{low, \tau}\equal 0$ for $E\in\{Y, D\}$.  We assign treatment based on the propensity score $\pi(x)\equal\text{expit}(\xi(x_{S_A} \minus 0.5))$ where both $\xi \!\in \! [-6,6]$ and whether $S_A$ overlaps $S_E$ determines the strength of confounding. We generate samples of size $n\equal5000$ for $K\equal30$ time-steps, and elaborate further on the experiments and data generating processes (DGP) in Appendix \ref{app:details}\footnote{Code to replicate all experiments can be found at \url{https://github.com/AliciaCurth/CompCATE} or \url{https://github.com/vanderschaarlab/CompCATE}.}. Using this DGP, we consider three main settings:
\begin{enumerate}[noitemsep, leftmargin=*]
    \item \textbf{Setting 1:} There is confounding as $S_A=1$ when $|\xi|>0$, but treatment has no effect on either event. Varying $\xi$ should lead to different levels of covariate shift due to confounding for all effects.
    \item \textbf{Setting 2:} There is no confounding ($\xi=0$), treatment has no effect on the main event but affects the competing event ($p^D_{low, \tau}=.01$). Varying  $p^D_{high, \tau}$, the heterogeneous effect of treatment on competing events in the high-risk group, should lead to a covariate shift in the at-risk group when interventions associated with direct and separable effect are considered. 
    \item \textbf{Setting 3:} There is no confounding ($\xi=0$), treatment has a heterogeneous effect on the main event (it equalizes main event risk between both groups, $p^Y_{high, \tau}=-.09$) but has no effect on the competing event. When the high risk group is also at higher (baseline) risk of the competing event (as $p^D_{high}$ varies) this may mask the protective effect of treatment on the main event, and should lead to a covariate shift in the at-risk group for the intervention associated with the direct effect only. 
\end{enumerate}

{\textbf{Estimators. } We focus on a setup where $\hat{h}_Y(k, x, a)$ is \textit{misspecified}, illustrating the effects of covariate shift on the main outcome model. We assume that the competing event model can be correctly specified, which may be the case in reality if $D_k$ is a well-studied comorbidity. Due to the simple DGP, both models could easily be estimated with an (unrestricted) logistic regression (LR) per time step using the classification framework discussed in the previous section. We use unrestricted (and hence correctly specified) LRs for $h_D(k, x, a)$, but induce misspecfication in $h_Y(k, x, a)$ by fitting \textit{constant} $\hat{h}^Y_{a, k}$ for each $k\!\leq \!K$ (this is equivalent to fitting a simple Kaplan-Meier estimator \citep{kaplan1958nonparametric} by treatment arm). We also demonstrate that similar conclusions apply when using a LR with L2-penalty that is set too aggressively. 
To highlight how different covariate shifts affect learning of the different effects, we compare the estimates obtained by \textit{Observational} ERM to importance-weighted ERM with \textit{Estimated} weights $\hat{w}^{*, int}_{a, k}$ and to two oracle solutions: weighted ERM with \textit{true} importance weights ${w}^{*, int}_{a, k}$ and unweighted ERM on a \textit{counterfactual} sample of size $n$ from the (usually inaccessible) interventional distribution.}

\textbf{Evaluation Metrics. } Using independent test sets of size $n_{te}\equal10^4$, we report the root-mean-squared-error 
$RMSE_\tau\!\!=\!\!RMSE(\tau(x))$ of estimating the three different types of risk differences of Section \ref{sec:riskdefs} (capturing a total, direct and separable direct HTE), which corresponds to an adaptation of \cite{hill2011bayesian}'s popular Precision in Estimating Heterogeneous Effects (PEHE) metric to our setting. To link back to our theoretical analysis, we also report the RMSE of estimating the hazard function, $RMSE^{a,k}_{haz}\equal \sqrt{\mathbb{E}_{X \sim \mathbb{P}^{int}_{a, k}}[(h^Y(l, X, a)\minus\hat{h}^Y(l, X, a))^2]}$ where $P^{int}_{a, k}$ is the interventional at-risk distribution corresponding to the effect of interest. We report mean and standard error across 10 replications of each experiment.

\subsection{Empirical insights}
\textbf{\textit{\textbullet{} Confounding-induced covariate shifts indeed impact estimation of all effects. }} In Fig. \ref{fig:conf_res} we present results for all 3 settings, highlighting that some effect estimates are indeed impacted by \textit{more} covariate shifts than others. In first setting, we observe that \textit{all effect estimates} are impacted by increasing confounding strength $|\xi|$: standard ERM performs poorly while importance weighting performs almost identically to the counterfactual solution.

\textbf{\textit{\textbullet{} Shifts induced by competing events indeed do not affect estimation of all effects.}}
Next, we consider settings 2 and 3 in which treatment assignment is random, but covariate shift can arise due to a differential effect of treatment on the competing event (Setting 2), or due to the high risk group also being at higher risk of the competing event which may mask a protective treatment effect on the main event (Setting 3). In Fig. \ref{fig:conf_res} we observe that, as expected, bias due to covariate shift in setting 2 arises only for separable and direct effect, both of which require elimination of the differential effect of treatment on the competing event -- while the total effect does not. Setting 3 exhibits a covariate shift that induces bias only for the direct effect as expected, as treatment has no separable indirect effect here. We also  find that, while the error in estimation of $h^Y_{a, k}$ appears small at each time step, it becomes substantial as all $K$ separate hazard functions are cumulated and used to estimate the difference in risk. Note that the absolute error in estimation of the hazard appears to \textit{decrease} over time in most settings -- this may appear counterintuitive at first glance as $n^Y_{a,k}$ decreases over time, but is expected in our setting where the interventional at-risk population becomes more homogeneous over time as only low risk individuals are expected to survive, making the constant model approximately correct for later time steps. 
\begin{figure}[!b]
\vspace{-1em}
	\centering
	\subfigure[ESS in setting 2.]{\label{fig:ess}\includegraphics[width=0.48\columnwidth]{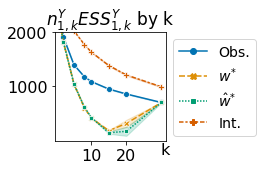}}
    \subfigure[$RMSE_\tau$ in setting 4.]{\label{fig:set4}\includegraphics[width=0.48\columnwidth]{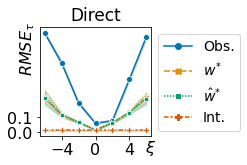}}
\vspace{-1em}
	\caption{Additional results. Left: effective sample size (ESS) for direct effect estimation in setting 2 ($p^D_{high, \tau}\equal.2$). Right: $RMSE_\tau$ of direct effect estimation for varying confounding $\xi$ in presence of side effects (additional setting 4)}\label{fig:additional}   
\end{figure}

\textbf{\textit{\textbullet{} Effective sample size behaves and contributes as expected.}} In settings 2 \& 3, when considering $RMSE^{a,k}_{haz}$ across $k$, we observe a tradeoff as expected from our theoretical analysis: as $k$ increases, covariate shift becomes more extreme, initially widening the performance gap between weighted and unweighted ERM solution. As $k$ grows larger, the increased variance (low ESS) of the weighted solution then starts to hinder its performance. Finally, for very large $k$ all solutions converge as the target distribution becomes more homogeneous. This tradeoff is indeed also reflected in the (absolute) effective sample sizes as measured by $\textstyle ({\sum_{i \in \mathcal{I}^Y_{a,k}}\bar{w}^{*,2}_i})^{-1}$ in Fig. \ref{fig:ess}.

\textbf{\textit{\textbullet{} Multiple shifts can offset or exacerbate each other. }} Finally, we combine settings 1 and 2 (setting 4: $p^D_{high, \tau}=0.1$, $\xi$ varies); thus the high risk group experiences adverse reactions to treatment in the competing event, and setting $\xi>0$ ($\xi<0$) corresponds to assigning more (less) high risk individuals to treatment. In Fig. \ref{fig:set4}, we observe that having assigned more high risk individuals to treatment can offset the shift induced by competing events for (separable) direct effects (and, conversely, the more sensible practice of assigning less high risk individuals to treatment would exacerbate the covariate shift for (separable) direct effects).

\begin{figure}[h]
%\vskip -0.1in
    \centering
    \includegraphics[width=0.99\columnwidth]{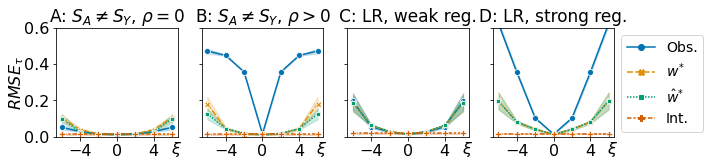}
    \vspace{-1em}
    \caption{Variations on setting 1 with constant estimators (A \& B) and logistic regressions (C \& D)}
 \label{fig:mattershifts}\vspace{-2em}
\end{figure}

\textbf{\textit{\textbullet{} When do such covariate shifts truly matter?}} In Fig. \ref{fig:mattershifts} we finally investigate \textit{when} shifts matter for ERM. We revisit setting 1, and, in panels A\&B, highlight that when $x_{S_A}$ is not a true confounder (does not affect $Y_k$), the resulting covariate shift biases ERM only when $x_{S_Y}$ also shifts due to correlation with $x_{S_A}$ (the same holds true for other shifts when ${S_Y}\!\neq \!S_D$). In panels C\&D, we confirm the impact of misspecification by using a LR as outcome model: an unrestricted LR (C) can fit the DGP well and covariate shift thus has little effect, while the introduction of excessive regularization (D) leads to the need to prioritize regions of $\mathcal{X}$ (which observational ERM does incorrectly). 

\section{\uppercase{Conclusion}} We studied the challenges inherent to HTE estimation in the presence of competing events, and found that inclusion of competing events not only leads to multiple definitions of effects but also to multiple sources of covariate shifts when estimating them. Theoretically and empirically, we highlighted \textit{that}, \textit{when} and \textit{how} different shifts bias estimation of different effects. Having gained understanding of its challenges, an interesting next step, further discussed in Appendix \ref{app:ext}, would be to consider how to adapt ideas from the ML literature on HTE \& TTE estimation -- such as the deep treatment-specific hazard estimator of \citet{curth2021survite} -- to construct more sophisticated solutions for our problem setting. 

Finally, note that we do \textit{not} wish to argue here that one measure of effect is superior to others -- instead, we only aim to  raise awareness that using different types of outcome predictors can lead to different effect interpretations and that the challenges inherent to learning them differ. Ultimately, in applications the choice of estimand should be made by domain expert -- who is also needed to verify that its \textit{untestable} identifying assumptions hold -- and correspond to their policy or research question of interest.  

\newpage
\subsubsection*{Acknowledgements} We would like to thank anonymous reviewers and members of the vanderschaar-lab for insightful comments on earlier drafts of this paper, and Mats Stensrud for a very interesting discussion on the topic. AC gratefully acknowledges funding from AstraZeneca.

\bibliographystyle{apalike}
\bibliography{main.bib}

\newpage
\onecolumn
\appendix
\section*{\uppercase{Appendix}}
This Appendix is structured as follows: In Appendix A, we present an additional literature review. In Appendix B, we discuss possible extensions incorporating more sophisticated solutions from the ML literatures on TTE prediction and HTE estimation. In Appendix C, we discuss identifying assumptions and proofs. In Appendix D, we give additional details of the simulation experiments presented in the main text. In Appendix E, we finally conduct additional experiments based on real data. 

\section{\uppercase{Additional literature review}}\label{app:lit}
\paragraph{Time-to-event prediction using ML in the presence of competing events.}  While modeling time-to-event data in the presence of competing events has been studied in the  statistics literature for decades \citep{prentice1978analysis, gray1988class, fine1999proportional}, 
\cite{lee2018deephit} are the first paper we are aware of that consider time-to-event analysis with competing events in a (deep) machine learning context. \cite{lee2018deephit}'s proposed DeepHit assumes a discrete-time setup and uses fully connected neural networks with both shared and time-specific components for prediction of cause-specific incidence at every time-step.   This work has since been extended further proposing improvements upon using simple neural networks in discrete time e.g. using RNNs in \cite{li2020deepcomp} or transformers in \cite{wang2022survtrace} and been complemented with work in continuous time relying on e.g. multi-task boosting \citep{bellot2018multitask} or Bayesian nonparametric methods e.g. multi-task gaussian processes \citep{alaa2017bayesian}, Lomax delegate racing \citep{zhang2018nonparametric} or tree-based mixture models \citep{bellot2018tree}.

\paragraph{Heterogeneous treatment effect estimation using ML. } 
The ML literature on HTE estimation has centered mainly on binary or continuous outcomes, and has expanded rapidly over the last years. One stream of work has provided model-agnostic strategies (also sometimes referred to as meta-learner strategies \citep{kunzel2019metalearners}) to estimate HTEs using \textit{any} ML method \citep{kunzel2019metalearners, nie2017quasi, kennedy2020optimal, curth2020}. A majority of the work published in machine learning has, however, focussed on adapting \textit{specific} ML methods for HTE estimation: early work relied mainly on tree-based methods \citep{hill2011bayesian, athey2016recursive, wager2018estimation, athey2019generalized}, but was followed by adaptations of e.g. Gaussian processes \cite{alaa2017bayesian, alaa2018limits} and GANs \cite{yoon2018ganite}. At this point, the most popular solution seems to be to adapt neural networks for treatment effect estimation, see e.g.  \cite{johansson2016learning, shalit2017estimating, johansson2018learning, shi2019adapting, hassanpour2019counterfactual, hassanpour2020learning, assaad2020counterfactual, curth2020, curth2021inductive}.  The work outlined above focusses exclusively on binary/continuous outcomes, thus closest to our setting are two recent papers that have investigated challenges inherent to HTE estimation for TTE data \textit{without} competing events, focussing on covariate shift: \cite{chapfuwa2021enabling} used generative models for counterfactual TTE analysis in continuous time and \cite{curth2021survite} used neural networks for discrete time analyses but neither considers how to incorporate competing events.

\paragraph{Estimating treatment effects in the presence of competing events.} The most likely reason for a lack of work on estimating \textit{heterogeneous} treatment effects in the presence of competing events is that even the simpler \textit{average} treatment effect setting has received rigorous characterisation within a causal framework only very recently: \cite{young2020causal, stensrud2020separable} are the first to formally characterize and define different types of causal effects and their identifying conditions within a counterfactual framework; their formalization allowed us to extrapolate their insights, combined with the literature on HTEs from the standard settings, to \textit{heterogeneous} effects. Lacking such unified characterisation, prior work has considered estimation of average effects either in a model-dependent fashion by considering regression coefficients in cause-specific hazard models \cite{prentice1978analysis} or by testing for treatment-differences across e.g. cause-specific cumulative incidence functions \cite{gray1988class}, thus considering mainly total effects.

\section{\uppercase{Possible methodological extensions}}\label{app:ext}
Because we put our focus on understanding the unique challenges in adding competing events to the HTE estimation problem, we relied on simple ML methods to allow for clear empirical insights. Having gained understanding of the  challenges inherent to the HTE competing events problem, an interesting next step would therefore be to consider how to adapt and incorporate ideas from the vast ML literature on HTE \& TTE estimation to construct more sophisticated solutions. We discuss some possible avenues of interest for future research below. 

A first approach would be to make the time-to-event (hazard) predictions -- which we use to compute the effects --  themselves better, for example by sharing information across time-steps. As described in section \ref{sec:riskdefs}, any time-to-event model that allows to compute cause-specific hazard functions for every time-step could be used to estimate all three types of effects. Instead of fitting separate models per time-step as we do in our experiments, one could therefore flexibly share information across time-steps (and possibly across causes) as in discrete-time neural networks for TTE prediction \citep{lee2018deephit, li2020deepcomp, curth2021survite, wang2022survtrace}. Complementing such a methodological extension, it would also be interesting to theoretically study whether sharing of information over time-steps might mitigate some covariate shift issues.

Further, the literature on domain adaptation and HTE estimation has proposed more sophisticated solutions to address covariate shift than classical importance weighting. As we allude to in footnote 4 in the main text, a large proportion of the ML HTE literature (e.g. \cite{shalit2017estimating, johansson2018learning, assaad2020counterfactual}) has focused on learning balanced representations that minimize an empirical estimate of the IPM-term -- which is not readily available in our setting as the interventional at-risk distribution is not equal to the marginal covariate distribution. Faced with a similar obstacle, \cite{curth2021survite} simply used the marginal distribution for balancing nonetheless -- which they demonstrated to work well empirically. Improving upon this naive solution by investigating how to correctly balance representations may be an interesting next step both in their and our setting. A different avenue would be to improve upon using standard importance weights by removing uninformative dimensions: if we knew which features caused outcome, we would only need to compute importance weights taking into account distributional differences in dimensions that actually matter for prediction, which could significantly reduce variance in the importance weights and hence speed up learning \citep{stojanov2019low, maia2022effective}. One possible way of doing so might be to jointly learning importance weights and representations for time-to-event prediction, adapting ideas from e.g. \cite{hassanpour2019counterfactual, fang2020rethinking}

Finally, an interesting future direction may be to consider more complex data-types, e.g. allowing for (some) patient characteristics to be repeatedly measured over time, necessitating the incorporation of time-varying covariates. This is also an interesting scenario because it makes identifying assumptions relying on measuring all common causes of $Y_k$ and $D_k$ (Assumptions D1 and S1 in the following section) more likely to hold. Such covariates could easily be incorporated in our problem formulation and possible solutions could rely on recurrent networks such as \cite{lee2019dynamic} in the TTE prediction setting and \cite{bica2020estimating} in the longitudinal treatment effect estimation setting.

\section{\uppercase{Assumptions and Proofs}}\label{app:tech}
\subsection{Formal Presentation of Identification Assumptions}
Below we discuss assumptions for nonparametric \textit{identification} of effects, which are based on those given in \cite{young2020causal} for total and direct effect and \cite{stensrud2020separable} for separable effects. 

Some assumptions are shared by all causal parameters described in Section 3; they are analogues to the standard \textit{ignorability} assumptions of \cite{rosenbaum1983central} from the standard treatment effect setting and are known as \textit{unconfoundedness}, \textit{overlap} and \textit{consistency} assumptions: for each $k\in\{1, \ldots, K\}$ we require \\
\textbf{\textbullet{ } Assumption 1. Exchangeability w.r.t. treatment: } $Y^a_k, D^a_k\indep A|X$ \\
\textbf{\textbullet{ } Assumption 2. Positivity w.r.t. treatment: } $\mathbb{P}(A=a|X=x)>0$ for $\forall x: \mathbb{P}(X=x)>0$ and $a\in \{0, 1\}$\\
\textbf{\textbullet{} Assumption 3. Consistency w.r.t. treatment assignment: } We observe the counterfactuals associated with the given treatment $A$, i.e. $Y_k=AY^1_k+ (1-A)Y^0_k$ and $D_k=AD^1_k+ (1-A)D^0_k$

While total effects require no additional assumptions, both direct and separable effects require further identification assumptions associated with the additional interventions performed in their definitions.

\paragraph{Direct effects.} Direct effects require additional unconfoundedness, overlap and consistency assumptions to identify distributions under elimination of the competing event. For each $k\in\{1, \ldots, K\}$ we require\\
\textbf{\textbullet{} Assumption D1. Exchangeability: w.r.t. competing event}  $Y^a_k \indep D^a_k|X, \bar{Y}_{k-1}=\bar{D}_{k-1}=0, A=a$ \\
\textbf{\textbullet{ }Assumption D2: Positivity w.r.t. competing event: } $\mathbb{P}(D_k=0|X=x, \bar{Y}_{k-1}=\bar{D}_{k-1}=0, A=a)>0$ whenever $\mathbb{P}(X=x, \bar{Y}_{k-1}=\bar{D}_{k-1}=0, A=a)>0$ \\
\textbf{\textbullet{} Assumption D3. Consistency w.r.t. elimination of competing event:} For an observation with $A=a$ and $\bar{D}_k=0$ we observe the corresponding counterfactual, i.e. $\bar{Y}_k=\bar{Y}_k^{a, \bar{d}=0}$. \\

\paragraph{Separable effects.} 
\cite{stensrud2020separable} derived assumptions enabling identification of separable (in)direct effects; owing to the conceptual difference in intervention on only \textit{parts} of the treatment, these are more involved to state than those above. In particular, for each $k\in\{1, \ldots, K\}$ we require: \\
\textbf{\textbullet{} Assumption S0. Conceptual assumptions defining separable treatment: } $A$ is separable into components $A_Y$ and $A_D$. Setting $A=a$ is equivalent to setting both $A_Y$ and $A_D$ to $a$, so that $Y^{a_Y=a, a_D=a}_k=Y^{a}_k$ and$D^{a_Y=a, a_D=a}_k=D^{a}_k$. Further, $A_Y$ exerts effects on $D_k$ only through its effect on $\bar{Y}_{k-1}$ i.e. \begin{equation}
    Y^{a_Y=1, a_D}_{k-1}= D^{a_Y=1, a_D}_{k-1}= Y^{a_Y=0, a_D}_{k-1}= D^{a_Y=0, a_D}_{k-1}= 0 \implies D^{a_Y=1, a_D}_{k}= D^{a_Y=0, a_D}_{k} \text{ for } a_D \in \{0, 1\}
\end{equation} and conversely,  $A_D$ exerts effects on $Y_k$ only through its effect on $\bar{D}_{k}$ i.e. \begin{equation}
    Y^{a_Y, a_D=1}_{k-1}= D^{a_Y, a_D=1}_{k}= Y^{a_Y, a_D=0}_{k-1}= D^{a_Y, a_D=0}_{k}= 0 \implies Y^{a_Y, a_D=1}_{k}= Y^{a_Y, a_D=1}_{k} \text{ for } a_Y \in \{0, 1\}
\end{equation}\\
\textbf{\textbullet{ }Assumption S1. Dismissible Component conditions:} W.r.t. primary event \begin{equation}\begin{split}
\mathbb{P}(Y^{a_Y, a_D=1}_k=1|Y^{a_Y, a_D=1}_{k-1}=0, D^{a_Y, a_D=1}_{k}=0, X=x)=\\ \mathbb{P}(Y^{a_Y, a_D=0}_k=1|Y^{a_Y, a_D=0}_{k-1}=0, D^{a_Y, a_D=0}_{k}=0, X=x)
\end{split}
\end{equation} and w.r.t. competing event\begin{equation}\begin{split}
\mathbb{P}(D^{a_Y=1, a_D}_k=1|Y^{a_Y=1, a_D}_{k-1}=0, D^{a_Y=1, a_D}_{k}=0, X=x)=\\ \mathbb{P}(D^{a_Y=0, a_D}_k=1|Y^{a_Y=0, a_D}_{k-1}=0, D^{a_Y=0, a_D}_{k}=0, X=x)
\end{split}
\end{equation}
\textbf{\textbullet{ }Assumption S3. Positivity w.r.t. treatment (in surviving population): } $\mathbb{P}(A=a|\bar{D}_k=\bar{Y}_k=0, X=x)>0$ whenever $\mathbb{P}(\bar{D}_k=\bar{Y}_k=0, X=x)>0$ for $a\in\{0,1\}$\\
\textbf{\textbullet{} Assumption S3. Consistency:} For an observation with $A=a$, we observe the corresponding counterfactuals, i.e. $Y_k=Y^{a, a}_k$ and $D_k=D^{a, a}_k$.
\\

\subsection{Discussion of Assumptions. } Consistency assumptions are always needed to ensure that we  observe \textit{one} of the counterfactuals for each individual; these assumptions may not hold if e.g. the act of monitoring outcomes can change their value or if there is interference between individual units. Positivity assumptions ensure that there is some overlap between observational and interventional distributions; if these assumptions do not hold we could not (nonparametrically) extrapolate to the interventional distribution (and importance weights $w^*(x)$ would not be defined for all $x$). 

\begin{figure}
    \centering
    \includegraphics[width=.99\textwidth]{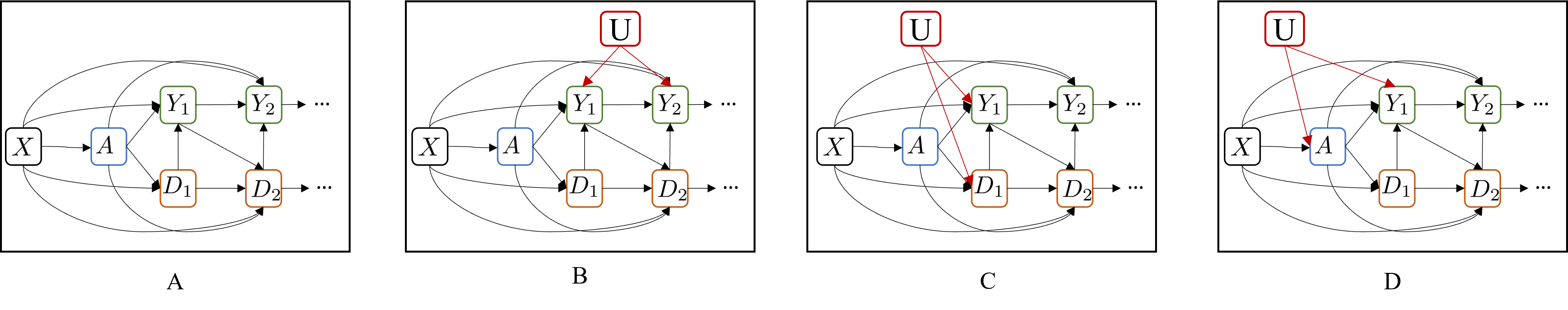}
    \caption{Figure illustrating causal graphs with hidden variables in which different effects are identified. (A): No hidden variables, all effects are identified. (B): Hidden variable causing all $Y_k$, all effects are identified. (C): Hidden variable causing $Y_k$ and $D_k$; total effect is identified but separable and direct effect are not. (D): Hidden confounder of outcome treatment association, no effect is identified. }
    \label{fig:id}
\end{figure}

Finally, exchangeability assumptions and the dismissible component assumptions ensure that there are no \textit{hidden} variables (variables not included in $X$) causing treatment assignment \textit{and} events (all effects) and $Y_k$ \textit{and} $D_k$ (separable and direct effect only), which would make observed distributions inherently different from distributions under intervention. To illustrate when these assumptions do not hold, we highlight scenarios where different hidden variables do (not) violate assumptions in Fig. \ref{fig:id}. In Panel A (the same as Fig. 1 in the main text), no hidden variables exist and all effects are identified. In Panel B, hidden variables causing both $Y_1$ and $Y_2$ exist -- e.g. some form of underlying frailty \citep{aalen1994effects}, inducing heterogeneity in risk of \textit{only} the main event occuring -- which are allowed under all effects. In Panel C, $Y_k$ and $D_k$ are caused by some shared hidden variable, which violates assumptions D1 and S1 -- thus separable and direct effects are not identified, while the total effect is. Finally, in Panel D, there is a hidden confounder of the treatment-outcome association, which violates assumption 1 -- therefore, no effect is identified. 

\subsection{Identification of interventional at-risk covariate distribution for separable effects (Sec. 4.2) }
The interventional at-risk covariate distribution corresponding to the intervention $do(A_Y=a_Y, A_D=a_D)$ can be identified from observational data under the assumptions stated above and is proportional to $\mathbb{P}(X \equal x)
       (1 \minus h_D(k, x, a_D)) \times \prod^{k-1}_{l=1}  (1\minus h_D(l, x, a_D))(1\minus h_Y(l, x, a_Y))$ as stated in Sec. 4.2. \\
\textbf{Proof:} the interventional distribution is proportional to 
\begin{equation}
\begin{split}
    \mathbb{P}(X\equal x) \mathbb{P}(D^{a_Y, a_D}_k \equal 0| D^{a_Y, a_D}_{k-1}\equal 0, Y^{a_Y, a_D}_{k-1}\equal 0, X\equal x) \\ \times \prod^{k-1}_{l=1} \mathbb{P}(Y^{a_Y, a_D}_{l} \equal 0| D^{a_Y, a_D}_{l}\equal 0, Y^{a_Y, a_D}_{l-1}\equal 0, X\equal x) \mathbb{P}(D^{a_Y, a_D}_l \equal 0| D^{a_Y, a_D}_{l-1}\equal 0, Y^{a_Y, a_D}_{l-1}\equal 0, X=x) 
\end{split}
\end{equation}
which is equal to 
\begin{equation}
\begin{split}
        \mathbb{P}(X\equal x) \mathbb{P}(D^{a_D, a_D}_k \equal 0| D^{a_D, a_D}_{k-1}\equal 0, Y^{a_D, a_D}_{k-1}\equal 0, X\equal x) \\ \times \prod^{k-1}_{l=1} \mathbb{P}(Y^{a_Y, a_Y}_{l} \equal 0| D^{a_Y, a_Y}_{l}\equal 0, Y^{a_Y, a_Y}_{l-1}\equal 0, X\equal x) \mathbb{P}(D^{a_D, a_D}_l \equal 0| D^{a_D, a_D}_{l-1}\equal 0, Y^{a_D, a_D}_{l-1}\equal 0, X=x) 
\end{split}
\end{equation}
because of the dismissible component conditions if $a_Y\neq a_D$, and trivially if $a_Y=a_D$.

This is equal to 
\begin{equation}
\begin{split}
       \mathbb{P}(X\equal x) \mathbb{P}(D^{a_D}_k \equal 0| D^{a_D}_{k-1}\equal 0, Y^{a_D}_{k-1}\equal 0, X\equal x) \\ \times \prod^{k-1}_{l=1} \mathbb{P}(Y^{a_Y}_{l} \equal 0| D^{a_Y}_{l}\equal 0, Y^{a_Y}_{l-1}\equal 0, X\equal x) \mathbb{P}(D^{a_D}_l \equal 0| D^{a_D}_{l-1}\equal 0, Y^{a_D}_{l-1}\equal 0, X=x) 
\end{split}
\end{equation}
\begin{equation}
\begin{split}
    = \mathbb{P}(X\equal x)  \mathbb{P}(D_k \equal 0| D_{k-1}\equal 0, Y_{k-1}\equal 0, X\equal x, A\equal a_D) \\ \times \prod^{k-1}_{l=1} \mathbb{P}(Y_{l} \equal 0| D_{l}\equal 0, Y_{l-1}\equal 0, X\equal x,  A\equal a_Y) \mathbb{P}(D_l \equal 0| D_{l-1}\equal 0, Y_{l-1}\equal 0, X=x,  A\equal a_D) 
    \end{split}
\end{equation}
\begin{equation}
    = \mathbb{P}(X \equal x)
       (1 \minus h_D(k, x, a_D)) \times \prod^{k-1}_{l=1}  (1\minus h_D(l, x, a_D))(1\minus h_Y(l, x, a_Y)
\end{equation}
by consistency (assumptions S0 and S3), exchangeability (assumption 1) and by definition of the hazard functions, respectively.

\subsection{Proof of proposition 1 (Section 4.3) }
\begin{proposition} (Restated) Given timestep $k$ and treatment $a$, for a loss function $\ell_h \in [0, 1]$ of any hypothesis $h \in \mathcal{H}$, such that $d=\text{Pdim}(\{\ell_h: h \in \mathcal{H}\})$ (where Pdim is the pseudo-dimension) and $\ell_h \in \mathcal{L}$, where $\mathcal{L}$ is a space of pointwise loss functions, and a weighting function $w(x)$ with $\mathbb{E}[w(X)]=1$, with probability $1 - \delta$ over at-risk sample $\mathcal{D}^Y_{a, k}$ with empirical distribution $\hat{p}^{obs}_{a, k}$, we have 
\vspace{-0.9em}
\begin{equation}\label{eq:thmmain}
\begin{split}
\textstyle R^{int}_{a, k}(h) - \hat{R}^{w, obs}_{a, k}(h) \leq \left |\mathbb{E}_{p^{obs}_{a, k}}[(w^*_{a, k}(x) - w(x))\ell_h(x)]\right| + 
\frac{\max(\sqrt{\mathbb{E}_{p^{obs}_{a, k}}[w^2(x)l^2_h(x)]}, \sqrt{\mathbb{E}_{\hat{p}^{obs}_{a, k}}[w^2(x)l^2_h(x)]})}{n^Y_{a, k}{}^{3/8}} \mathcal{C}^\mathcal{H}_{n^Y_{a, k}}
\end{split}
\end{equation}
with $\textstyle\mathcal{C}^\mathcal{H}_{n^Y_{a, k}} = 2^{5/4}(d \log \frac{e2n^Y_{a, k}}{d} + \log \frac{4}{\delta})^{3/8}$ due to \cite{cortes2010learning} (Theorem 4).

For $w(x)\equal w^*_{a, k}(x)$ (exact importance weights, Lemma 1), we have
\begin{equation}\label{eq:withweight}
     R^{int}_{a, k}(h) \minus \hat{R}^{w^*, obs}_{a, k}(h) \leq \frac{1}{\sqrt{ESS^{*}_{rel}(p^{int}_{a, k}, p^{obs}_{a, k})} n^Y_{a,k}{}^{3/8}} \mathcal{C}^\mathcal{H}_{n^Y_{a, k}}
\end{equation}

where $ESS^{*}_{rel}=\exp_2(D_2(p^{int}_{a, k}||p^{obs}_{a, k}))$ is the expected relative effective sample size, with $D_2(p||q)= log_2 \mathbb{E}_{x \sim p} \left[\frac{p(x)}{q(x)}\right]$ the R\a'enyi divergence of order 2.

For $w(x)=1$ (standard supervised learning, Lemma 2), we have 
\begin{equation}\label{eq:noweight}
    R^{int}_{a, k}(h) \minus \hat{R}^{obs}_{a, k}(h) \leq C_\mathcal{L} IPM_{\mathcal{L}}(p^{int}_{a, k}, p^{obs}_{a, k}) + n^Y_{a, k}{}^{-3/8} \mathcal{C}^\mathcal{H}_{n_{a, k}}
\end{equation}
where $\IPM_\mathcal{G}(p, q) = \sup_{g\in \mathcal{G}}\left|\int g(x) (p(x) - q(x))dx\right|$ is an integral probability metric and $C_\mathcal{L}\!>\!0$ is s.t. $\frac{\ell}{C_\mathcal{L}}\in \mathcal{L}$.
\end{proposition}

\paragraph{Proof: }
Eq. (\ref{eq:thmmain}) follows directly from Thm 4 in \cite{cortes2010learning}. 

Further, when $w(x)=w^*_{a, k}(x)$, the statement in eq. (\ref{eq:withweight},  follows directly from Theorem 3 of \cite{cortes2010learning}, where we used the restatement in terms of $ESS^{*}_{rel}$ of \cite{maia2022effective}.

Finally, to prove equation (\ref{eq:noweight}), note that when  $w(x)=1$, we can bound the first term of eq. (\ref{eq:thmmain}) as
\begin{equation}
\begin{split}
  \left|\mathbb{E}_{p^{obs}_{a, k}}[(w^*_{a, k}(x) - 1)\ell_h(x)]\right|= \left |\mathbb{E}_{p^{obs}_{a, k}}[(\frac{p^{int}_{a, k}}{p^{obs}_{a, k}} - 1)\ell_h(x)]\right| = \left|  \int  \ell_h(x) (p^{int}_{a, k}(x) - p^{obs}_{a, k}(x))dx\right| \\ \leq C_\mathcal{L} \sup_{f \in \mathcal{L}}  \left| \int f_h(x) (p^{int}_{a, k}(x) - p^{obs}_{a, k}(x))dx \right| \leq C_\mathcal{L} IPM_{\mathcal{L}}(p^{int}_{a, k}, p^{obs}_{a, k})
   \end{split}
\end{equation}
as in e.g. \cite{shalit2017estimating, johansson2018learning}'s bounds for the standard treatment effect estimation setting. Note further that because $\ell_h\in [0,1]$ by assumption, $\max(\sqrt{\mathbb{E}_{p^{obs}_{a, k}}[w^2(x)l^2_h(x)]}, \sqrt{\mathbb{E}_{\hat{p}^{obs}_{a, k}}w^2(x)l^2_h(x)]}) \leq 1$ in the second term. Putting the two together gives  (\ref{eq:noweight}).

\section{\uppercase{Experimental details}}\label{app:details}
\paragraph{Synthetic DGPs}
For clarity, we explicitly write out the DGPs used to create settings 1-4 in the main text below.
\begin{equation*}
\textbf{Setting 1 ($\xi$ varies): }  h_Y(k, x, a) = 0.01(1-x_1) + 0.1 x_1 \text{ and } h_D(k, x, a) = 0.01 \text{ and } \pi(x)\equal\text{expit}(\xi(x_1 \minus 0.5))
\end{equation*}
\begin{equation*}
\textbf{Setting 2 ($p^D_{high, \tau}$ varies): }  h_Y(k, x, a) = 0.01(1-x_1) + 0.1 x_1 \text{ and } h_D(k, x, a) = 0.01 + p^D_{high, \tau}*x_1*a \text{ and } \pi(x)=0.5
\end{equation*}
\begin{equation*}
\begin{split}
\textbf{Setting 3 ($p^D_{high}$ varies): }  h_Y(k, x, a) = 0.01(1-x_1) + (0.1 - 0.09a)*x_1 \text{ and } h_D(k, x, a) = 0.01(1-x_1) + p^D_{high}*x_1 \\\text{ and } \pi(x)=0.5
\end{split}
\end{equation*}
\begin{equation*}
    \textbf{Setting 4  ($\xi$ varies): }  h_Y(k, x, a) = 0.01(1-x_1) + 0.1 x_1 \text{ and } h_D(k, x, a) = 0.01 + 0.1*x_1*a\text{ and } \pi(x)\equal\text{expit}(\xi(x_1 \minus 0.5))
\end{equation*}
with $X_1 \!\sim\! \mathcal{B}(0.5)$ and $X_2 \sim \mathcal{B}(0.5 \minus \rho (1 \minus 2X_1))$ and $\rho=.35$ for all Figures except in Fig. 6 panel A\&B where $\pi(x)$ depends on $x_2$ instead of $x_1$.

%\textcolor{red}{If time: show histograms of distribution of the two groups over time!}

\textbf{Implementations\footnote{Code to replicate all experiments can be found at \url{https://github.com/AliciaCurth/CompCATE} or \url{https://github.com/vanderschaarlab/CompCATE}.}}
Throughout, as unrestricted/correctly specified estimators for the hazard of the competing event $\hat{h}_D$ for each time step and for all propensity estimators $\hat{\pi}$, we use logistic regressions (LR), relying on the sklearn\citep{scikit-learn} implementation with default parameters and  l2-penalty $C=100$. As described in the main text, we use constants to induce a misspecified hazard model for the main event at each time step . For the results using LRs for the main event hazard in Fig. \ref{fig:mattershifts}C\&D in the main text, we set l2-penalty $C=1$ in Fig. \ref{fig:mattershifts}C to reduce capacity of the model and create a misspecified model, and $C=100$ for the unrestricted version in Fig. \ref{fig:mattershifts}D.

\section{\uppercase{Additional experiments using real data}}\label{app:twins}
\subsection{Creating a semi-synthetic benchmark from the Twins dataset}
Due to the usual absence of counterfactuals in practice, benchmarking treatment effect estimation methods on real data is a challenging problem \citep{curth2021doing}. In our setting, this problem is exacerbated relative to the standard HTE setting because there are more interventions than 'just' intervention on treatment, meaning that there are even more unobserved counterfactuals (namely those corresponding to interventions on competing events or separable interventions). 

\textbf{The Twins Dataset. }The (real-world) Twins benchmark dataset used in \cite{louizos2017causal, yoon2018ganite} for binary outcomes and in \cite{curth2021survite} for TTE outcomes \textit{without} competing events is an interesting exception as Twins could be treated as their respective counterfactual under treatment -- which has been exploited in the standard setting where we only wish to intervene on treatment: The dataset consists of 11400 pairs of twins, for whom one can create a binary treatment such that $a=1$ denotes being the heaver twin at birth, and use this to emulate a hypothetical study measuring the HTE of birthweight on 1-year infant mortality (binary outcome) or survival times (in days, administratively censored at t=365). Note that, fortunately, the mortality rate is relatively low, giving an event rate of around 16\% over the full horizon. The dataset as used in \cite{yoon2018ganite, louizos2017causal, curth2021survite} contains 30 covariates for each twin relating to the parents, the pregnancy, and the birth (e.g., marital status, race, residence, number of previous births, pregnancy risk factors, quality of care during pregnancy, and number of gestation weeks prior to birth), of which we use 27 in our experiments (we dropped the three categorical features due to low variance).

\textbf{Semi-synthetic benchmarking setup. } We use this dataset as a basis for a semi-synthetic benchmarking setup in which we use the original covariates and event times to create an observational time-to-event dataset with competing events by (i) selectively observing only one twin during training and (ii) introducing a simulated competing event.  As both are simulated, we can intervene on these processes to give oracle solutions based on interventional distributions to compare to the solutions obtained from observational distributions. Note that, because most events happen early on -- 80\% of events happen on days 0-10, with 60\% of events occuring on day 0 -- we consider only the first 10 days, so that here $\mathcal{K}=\{0, \ldots, 10\}$.

In training, we selectively observe only one of the two twins, and for (i) induce confounding by sampling $A|X \sim \mathcal{B}(\pi(X))$ with propensity score $\pi(X_i)=\text{expit}(\xi_A \mathcal{Z}_{train}(|\mathcal{S}|^{-1}\sum_{p \in \mathcal{S}} X_{i,p}))$ where $\mathcal{Z}_{train}(\cdot)$ denotes standardization over the training set, $\xi_A$ determines the strength of selection and $\mathcal{S}$ is a feature subset which is chosen as discussed below. For (ii), using the observed trajectories $\bar{Y}^a_K$ from the Twins dataset, we introduce competing events by simulating $D^a_k \sim h_d(k, x,a )$ for units with $Y^a_{k-1}=0$, and setting all future values of $Y^a_k$  to zero whenever a competing event occurs. As a hazard we use
\begin{equation}
    h_D(k, x, a) = \begin{cases}
        \text{expit}\left(\log(0.1) + \xi_D (1-a) \mathcal{Z}_{train}(|\mathcal{S}|^{-1}\sum_{p \in \mathcal{S}} x_{p})\right) \text{ if } k=1\\
        \frac{0.1}{k-1} \text{ otherwise}
    \end{cases}
\end{equation}
which mimics the main outcome in that most events and heterogeneity occurs initially in period 0 and then levels of. As above  $\mathcal{Z}_{train}(\cdot)$ denotes standardization over the training set and $\xi_D$ determines the heterogeneity in the competing event process. Note that treatment here \textit{equalizes} the odds of the competing event across individuals. 

As before, we consider 3 settings to be able to disentangle the different forces at play here. In setting 1, we remove the competing event and consider only the effect of treatment selection by varying $\xi_A$. In setting 2, we assign treatment randomly and consider only the effect of the covariate shift induced by the competing event by varying $\xi_D$. In setting 3, we combine the two, set $\xi_A=2$ and vary $\xi_D$.

For both propensity and competing hazard, to ensure that the variables determining treatment assignment and competing event are actually important for the main event, we choose set $\mathcal{S}$ by selecting the most important covariates from a (treatment-agnostic) random forest for predicting mortality at time step 0 on the Twins data (using sklearn's \texttt{SelectFromModel} class). In Fig. \ref{fig:noov} in the experiments below, we show that choosing outcome-relevant features in such manner makes a difference in an experiment where we instead randomly sample a feature set of size $|\mathcal{S}|$.

\paragraph{Estimators.} Because the event data is real and not simulated, unlike in the main text, we do not know what correctly specified model is in this case. Here, we therefore consider as base models a constant model as in the main text, as well as a LR with cross-validated l1-penalty in $\{10^{-3}, 10^{-2}, 10^{-1}\}$ and a random forest with 100 trees. The competing events and propensity models are once more fit using a (correctly specified) LR with cross-validated l1-penalty in $\{10^{-3}, 10^{-2}, 10^{-1}, 1\}$. The four learning strategies -- observational, weighting (true and estimated) and counterfactual -- are as before. Note that because treatment assignment and competing event are simulated, ground truth weights and counterfactuals are accessible.

\paragraph{Evaluation.} As the main event data is real and ground truth probabilities are unknown, we instead evaluate all models in terms of their predictions of event-free restricted mean survival time under intervention $RMST^{int}_K = min(T^{int}, K)$ which we can compute from the observed (twins) event times $T^{Y, twins}(a)$ and simulated competing event times $T^{D, sim}(a)$ (both are set to K+1 if event never occurs)  as
\begin{equation*}
   RMST^{int}_K = \begin{cases}
  min(T^{Y, twins}(a), T^{D, sim}(a), K) \text{ if } int = do(A=a) \\
  min(T^{Y, twins}(a), K) \text{ if } int = do(A=a, \bar{D}=0) \\
  min(T^{Y, twins}(a_Y), T^{D, sim}(a_D), K) \text{ if } int = do(A_Y=a_y, A_D=a_D)
   \end{cases} 
\end{equation*}
The expected value of the RMST is equal to the area under the event-free survival curve, which for the different interventions can be computed from the hazard functions as
\begin{equation*} \mathbb{E}[RMST^{int}_K] = 
    \begin{cases}
    1+ \sum^{K-1}_{l=1} \left[\sum^{l}_{q=1} (1-h_D(q, x, a)) (1-h_Y(q, x, a))\right]\text{ if } int = do(A=a)\\
    1+ \sum^{K-1}_{l=1} \left[\sum^{l}_{q=1} (1-h_Y(q, x, a))\right]\text{ if } int = do(A=a, \bar{D}=0)\\
    1+ \sum^{K-1}_{l=1} \left[\sum^{l}_{q=1} (1-h_D(q, x, a_D)) (1-h_Y(q, x, a_Y))\right]\text{ if } int = do(A_Y=a_y, A_D=a_D)\\
    \end{cases}
\end{equation*}
Below we will sometimes refer to RMST(a); this refers to the different versions of RMST evaluated for $A=a$ at time $K=10$ (only for the separable direct effect, RMST(0) corresponds to $a_Y=a_D=0$ and RMST(1) corresponds to $a_Y=1$, $a_D=0$). We report the RMSE of estimating RMST using the hazards from the different models, $\sqrt{\frac{1}{n_{test}}\sum^{n_{test}}_{i=1}(RMST^{int}_{K,i} - \hat{\mathbb{E}}[RMST^{int}_K])^2}$; here we split the data 50/50 for training and testing (by twin pairs), and report means and standard errors of the RMSE across 5 replications of each experiment. Code to replicate the experiments will be released upon acceptance of the paper.

\subsection{Results using the Twins data}
\begin{figure}[t]
    \centering
    \includegraphics[width=0.99\textwidth]{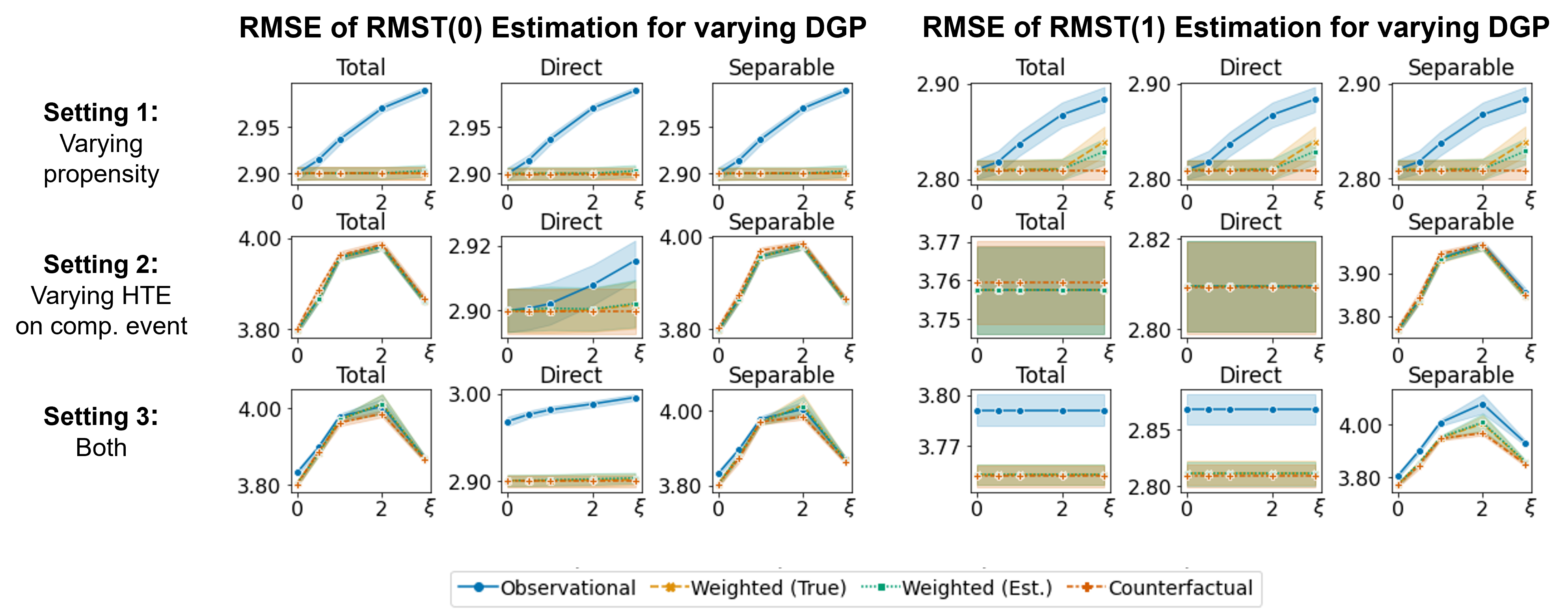}
    \caption{RMSE of estimating RMST under control (left) and under treatment (right) under the different interventions (columns) across different DGPs (rows) using (misspecified) constant estimators per time-step on Twins. Shaded area indicates one standard error.}
    \label{fig:twinscons}
\end{figure}
In Fig. \ref{fig:twinscons} we present results for all three settings and interventions corresponding to the three effects under consideration, using constant models at each time step to possibly induce misspecification. We observe that many of the conclusions from the stylized simulations from the main text carry over also to this more complex and realistic setup based on real data. We observe that varying confounding through $\xi_A$ (top row) remains important for all effects. We observe that only the estimator of the control\footnote{Note that, due to our hazard specification for the competing event, there is no covariate shift in the treatment group for the direct effect, thus it is expected that behaviour in estimation of RMST(1) is not impacted by $\xi_D$ in rows 2 and 3} RMST in the direct effect setting is impacted by varying the competing event intensity $\xi_D$ \textit{alone} (middle row), while when adding the two together we observe that varying $\xi_D$ can impact also estimation of the other effects.   

We also note that all differences between approaches overall appear much more salient for the RMST of the control group than for the treated group. While for settings 2 and 3 it is partially a consequence of how we designed the competing event hazard function, this is not the case for setting 1 -- here it may give some evidence that the outcome model in the treated population is not substantially misspecified and may be near constant; therefore we now focus on the control RMST for the remainder of this section. Additionally, we focus on estimation under intervention $do(A=a, \bar{D}=0)$, corresponding to direct effects, for which the results are most pronounced. 

\begin{wrapfigure}{r}{0.65\columnwidth}
\vskip -0.15in
    \centering
    \includegraphics[width=0.65\columnwidth]{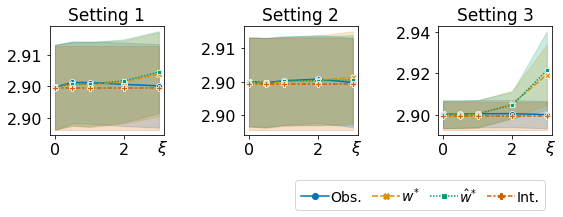}
    \vspace{-2em}
    \caption{RMSE of estimating control RMST under intervention $do(A=0, \bar{D}=0)$ (corresponding to direct effects) using a constant model for the three settings when features in $\mathcal{S}$ are randomly chosen.} \vspace{-5mm}
 \label{fig:noov}
\end{wrapfigure}We next consider the effect of having selected outcome-relevant covariates for treatment assignment and competing event. In Fig. \ref{fig:noov} we instead select random features and observe that indeed, compared to outcome-relevant features as in Fig. \ref{fig:twinscons}, the shift induced in the different settings no longer systematically plays a role.

\begin{figure*}[!h]
	\centering
	\subfigure[Random forest]{\label{fig:rf}\includegraphics[width=0.45\columnwidth]{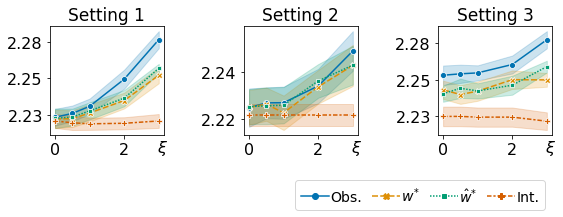}}
    \subfigure[Logistic Regression]{\label{fig:lr}\includegraphics[width=0.45\columnwidth]{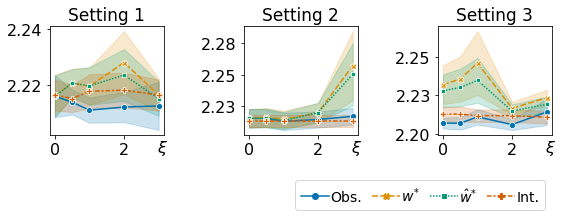}}

	\caption{RMSE of estimating control RMST under intervention $do(A=0, \bar{D}=0)$ (corresponding to direct effects) using random forests (left) and logistic regressions (right) for the three settings.}\label{fig:twinsml}   

\end{figure*}

Finally, we use more flexible models -- random forests (RFs) and logistic regressions (LRs) -- instead of the simple constant models (returning to a setting where $\mathcal{S}$ is outcome-relevant). In Fig. \ref{fig:twinsml} we present results for the three settings. We observe that for RFs, these covariate shifts appear to also play a role similarly to the constant estimator, albeit with smaller magnitude in effect on the estimation error. Interestingly, for LRs we do not observe any performance degradation of the observational solution relative to the counterfactual solution (and if anything, we observe that variance induced by weighting sometimes degrades performance). As robustness to arbitrary distribution shifts can indicate  correct specification \citep{wen2014robust}, this may provide some evidence that the LR-model is actually \textit{correctly} specified to capture the complexity of the underlying Twins time-to-event outcome data.

\end{document}